\documentclass[12pt]{article}

\usepackage{newtxtext,newtxmath}

\usepackage{graphicx}

\usepackage[letterpaper,margin=1in]{geometry}

\renewenvironment{abstract}
	{\quotation}
	{\endquotation}

\date{}

\makeatletter
\renewcommand{\fnum@figure}{\textbf{Figure \thefigure}}
\renewcommand{\fnum@table}{\textbf{Table \thetable}}
\makeatother

\usepackage{scicite}

\usepackage{url}

\def\scititle{{K$_2$Co$_2$(TeO$_{3}$)$_{3}$~$\cdot$~2.5~H$_2$O~:~A mineral-inspired pseudo-honeycomb cobalt dimer antiferromagnet}}
\title{\bfseries \boldmath \scititle}

\author{
	Austin~M.~Ferrenti$^{1,2,4,5\ast}$,
	Maxime~A.~Siegler$^{1}$,
	Yiqing~Hao$^{3}$,
        Chris~Lygouras$^{2}$,\and
        Tong~Chen$^{2}$,
        Tiffany~A.~Soetojo$^{2}$,
        Megan~R.~Rutherford$^{4,5}$,
        Kenji~M.~Kojima$^{7}$,\and
        Huibo~Cao$^{3}$,
        Natalia~Drichko$^{2}$,
        Alannah~M.~Hallas$^{4,5,6}$,
        Tyrel~M.~McQueen$^{1,2,8}$\and
	\small$^{1}$Department of Chemistry, The Johns Hopkins University, Baltimore, Maryland 21218, USA.\and
	\small$^{2}$Institute for Quantum Matter, William H. Miller III Department of Physics and Astronomy,\and \small The Johns Hopkins University, Baltimore, Maryland 21218, USA.\and
        \small$^{3}$Neutron Scattering Division, Oak Ridge National Laboratory, Oak Ridge, Tennessee 37831, USA\and
        \small$^{4}$Quantum Matter Institute, University of British Columbia, Vancouver, BC V6T 1Z4, Canada\and
        \small$^{5}$Department of Physics \& Astronomy, University of British Columbia, Vancouver, BC V6T 1Z1, Canada\and
        \small$^{6}$Canadian Institute for Advanced Research (CIFAR), Toronto, ON, M5G 1M1, Canada\and
        \small$^{7}$Centre for Molecular and Materials Science, TRIUMF, 4004 Wesbrook Mall, \and \small Vancouver, BC V6T 2A3, Canada \and
        \small$^{8}$Department of Materials Science and Engineering, The Johns Hopkins University, \and \small Baltimore, Maryland 21218, USA\and
	\small$^\ast$Corresponding author. Email: austin.ferrenti@ubc.ca\and
}

\begin{document} 

\maketitle

\begin{abstract} \bfseries \boldmath
 In recent years, magnetically-frustrated triangular and honeycomb lattice cobaltates have seen extensive study in the pursuit of a quantum spin liquid (QSL) state in a real material. In this work, we describe the hydroflux synthesis of K$_2$Co$_2$(TeO$_3$)$_3$ $\cdot$~2.5~H$_2$O (KCoTOH), a novel zemannite-type antiferromagnet (AFM) possessing structural elements of both triangular dimer and honeycomb structural motifs. Bulk magnetometry and specific heat data support the onset of long-range AFM order below $T_\text{N}$~=~7.6(1)~K, with neutron diffraction and muon spin relaxation ($\mu$SR) measurements placing the majority of the ordered moment within the pseudo-honeycomb plane. We resolve three unique oscillation frequencies from the zero-field $\mu$SR spectra, additionally suggesting a remarkably low level of structural disorder in as-grown KCoTOH crystals. Whereas interactions between dimerized chains of Co$^{2+}$ cations are typically observed to be negligible or ferromagnetic in nature, the largely planar ordering motif observed in KCoTOH is instead stabilized by net antiferromagnetic interactions through bridging tellurite groups. This work highlights the potential of hydroflux synthesis methods in the stabilization of magnetic materials possessing novel and potentially more frustrated lattice geometries.  
\end{abstract}

\noindent
The pursuit of new regimes of frustrated magnetism remains limited by the small number of unique lattice geometries known to host it. In the past decade, significant efforts have been devoted to the study of both spin-1/2 triangular and honeycomb lattice antiferromagnets, whose restricted geometry and highly anisotropic exchange frustrate neighboring magnetic moments.\cite{balents2010spin,broholm2020quantum} Several $S_\text{eff}$~=~1/2 triangular lattice systems, such as YbMgGaO$_4$ \cite{paddison2017continuous} and the cobalt dimer compound K$_2$Co(SeO$_3$)$_2$,\cite{zhong2020frustrated} as well as proposed Kitaev systems $\alpha$~-~RuCl$_3$,\cite{banerjee2017neutron} H$_3$LiIr$_2$O$_6$,\cite{halloran2025continuum} and Na$_2$Co$_2$TeO$_6$,\cite{miao2024persistent} have been shown to exhibit exotic spin dynamics down to the lowest accessible temperatures.\cite{chamorro2020chemistry} Despite this observation of strong quantum fluctuations, the majority of reported QSL candidate materials have been shown to ultimately order antiferromagnetically or exhibit spin freezing at low temperatures, motivating the exploration of novel structure types.

One potential testbed for further structural development is the family of honeycomb cobaltates, including Na$_2$Co$_2$TeO$_6$ and BaCo$_2$(AsO$_4$)$_2$, \cite{miao2024persistent,halloran2023geometrical} which were initially proposed as proximate realizations of the exactly solvable Kitaev quantum spin liquid (KQSL) model, defined by highly anisotropic exchange between Ising type spins on an otherwise unfrustrated lattice.\cite{kitaev2006anyons} While the Kitaev model assumes ideal octahedral coordination of the Co$^{2+}$ cation in an edge-sharing arrangement, significant distortions in real materials typically give rise to the strengthening of other non-Kitaev-type exchange between nearest neighbor and third nearest neighbor moments, complicating the identification of intrinsic Kitaev physics. In addition, recent theoretical and experimental studies have suggested that nearest neighbor Co-Co distances in such structures are simply too short to constrain direct exchange contributions from nearest neighbors in the ground state spin Hamiltonian.\cite{maksimov2022ab} Reducing the magnitude of this direct exchange would require greater distance between neighboring Co$^{2+}$ cations, while still remaining close enough to stabilize strong Kitaev-type exchange. 

Such a structure is partially realized in the naturally-occurring mineral zemannite,\\ Mg$_{0.5}$ZnFe$_3$(TeO$_3$)$_3$~$\cdot$~$x$~H$_2$O, where the Zn$^{2+}$~/~Fe$^{3+}$ cations form a buckled honeycomb lattice, with each cation well-separated in the \textit{ab}-plane by large polyanion groups, while Mg$^{2+}$ and H$_2$O fill the hexagonal channels. The cations forming the buckled honeycomb also dimerize along \textit{c}, producing an overall intermediate structure between that of the more commonly-studied triangular and honeycomb lattice systems. Several synthetic analogues of natural zemannite have been previously reported, \cite{missen2019crystal,huang2021crystal,eder2023structural} however their magnetism has rarely been studied. \cite{liu2022magnetism} Here we report the synthesis of zemannite-type antiferromagnet K$_2$Co$_2$(TeO$_{3}$)$_{3}$~$\cdot$~2.5~H$_2$O (hereafter referred to as KCoTOH), containing dual triangular lattices of [Co$_2$O$_9$] dimers slightly offset along $b$, producing an overall pseudo-honeycomb arrangement of Co$^{2+}$ cations separated by large [TeO$_3$]$^{2-}$ groups within the \textit{ac}-plane. The large honeycomb channels, similar to those observed in metal-organic framework and inorganic zeolite materials,\cite{armbruster2001crystal,furukawa2013chemistry} here accommodate chiral chains of potassium ions which break all rotational symmetries of the underlying cobalt tellurate framework. Despite the close proximity of the Co$^{2+}$ cations along the dimer-chain axis, our results suggest that KCoTOH does not display the significant magnetic anisotropy intrinsic to other dimerized triangular lattice antiferromagnets, but rather behaves more akin to a two-dimensional honeycomb system. Elastic neutron diffraction measurements provide evidence for a commensurate, antiferromagnetic ground state with a largely planar ordering motif, highlighting the potential of the zemannite-type structure to realize more complex, frustrated magnetic ground states.

\subsection*{Structural characterization}
\begin{figure}[ht]
    \centering
    \includegraphics[width=0.8\textwidth]{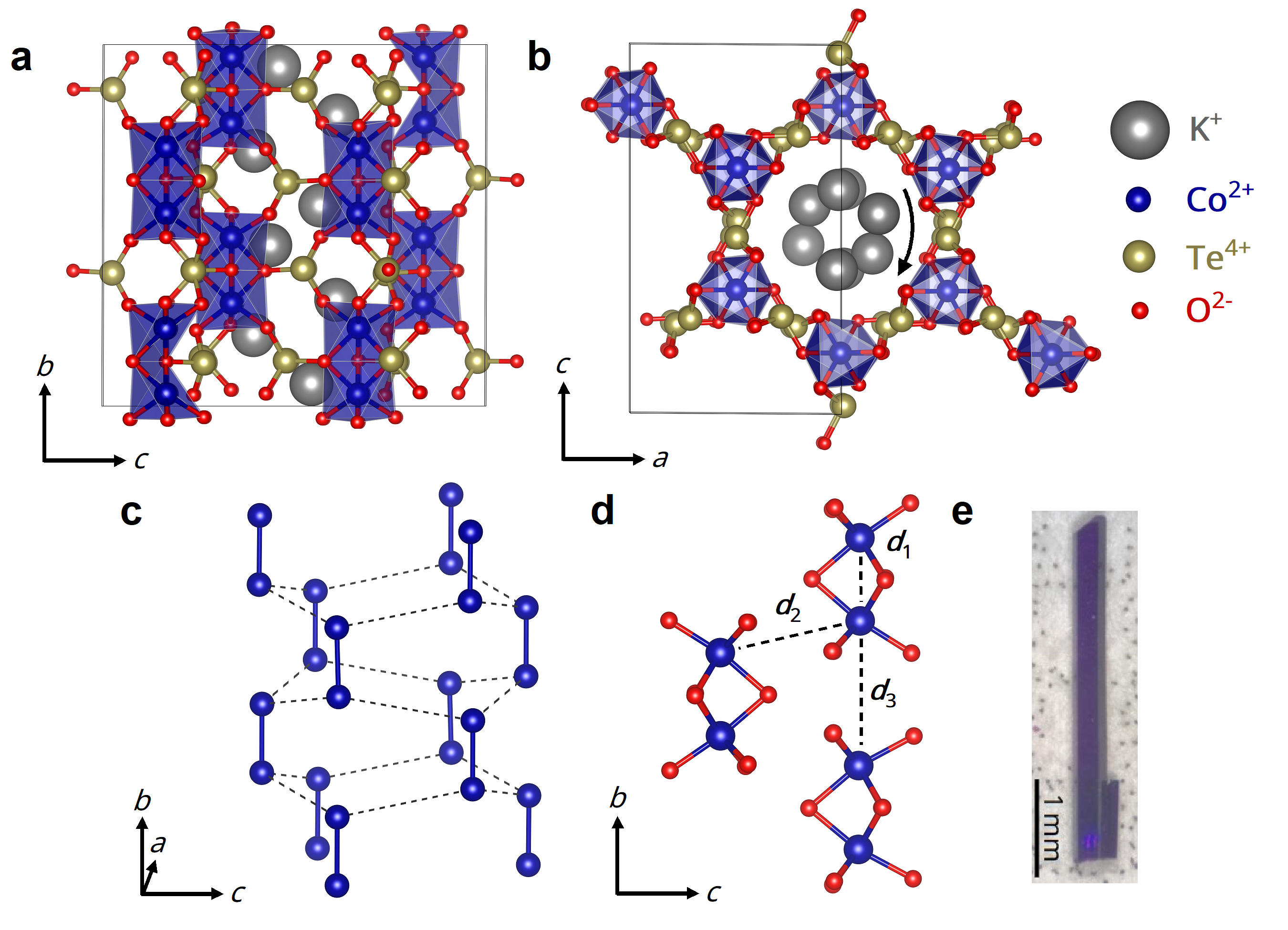}
    \vspace{-5mm}
    \caption{\textbf{Crystal structure of triangular dimer, pseudo-honeycomb K$_2$Co$_2$(TeO$_{3}$)$_{3}$~$\cdot$~2.5~H$_2$O (KCoTOH)} projected along the a) \textit{a}- and b) \textit{b}-axes. The arrow shows the direction of chirality of the K-O-K chains within the honeycomb channels. When viewed within the \textit{ac}-plane, the arrangement of Co$^{2+}$ cations resembles that of a distorted honeycomb lattice, highlighted by the dashed black lines in c). d) While possessing structural elements of both the triangular dimer and honeycomb lattice cobaltates, all intra- ($d_1$) and inter-dimer ($d_2$, $d_3$) distances within the zemannite-type KCoTOH structure are longer than is typical in either known family. e) Image of a representative KCoTOH single crystal composite.}
    \label{KCoTOH_Main_struc}
\end{figure}

Initial attempts to stabilize novel magnetic cobaltate phases via the hydroflux method were found to result in the formation of clusters of small (4 mm $\times$ 0.5 mm $\times$ 0.5 mm), purple, needle-like crystals. Single crystal X-ray diffraction was used to identify the phase as K$_2$Co$_2$(TeO$_{3}$)$_{3}$~$\cdot$~2.5~H$_2$O (KCoTOH). KCoTOH crystallizes in the three-dimensional zemannite-type structure, with dual triangular lattices of zigzag chains of Co$^{2+}$ dimers running along the \textit{b}-axis (Figure \ref{KCoTOH_Main_struc}a,b). These triangular lattices, akin to those observed in other triangular dimer cobaltate phases, are rotated 180° within the \textit{ac}-plane with respect to one another, forming an overall pseudo-honeycomb arrangement of Co$^{2+}$ cations when viewed along \textit{b} (Figure \ref{KCoTOH_Main_struc}c).\cite{zhong2020frustrated} This corrugated honeycomb lattice, formed from the top Co$^{2+}$ from one dimer set and the bottom Co$^{2+}$ of another, represents an effective intermediate between the triangular dimer and honeycomb cobaltate systems which have seen extensive study in recent years. KCoTOH crystals were also grown in a deuterated form, and will be referred to as KCoTOD when employed for characterization, as varying hydrogen and deuterium is not expected to affect chemical reactivity or magnetic interactions. 

Our structure solution is in general agreement with that described by Eder \textit{et. al.} in their report of a variety of novel zemannite-type phases grown by a similar hydroflux method, as well as that reported for the Se-analogue, albeit here with full ordering of the honeycomb channel contents.\cite{eder2023structural,wildner1993zemannite} While zemannite-type systems initially appear to possess an overall hexagonal crystal structure, the ordered contents of the honeycomb channels often break rotational and/or mirror symmetries. The larger crystals afforded by the hydroflux method described in this work consistently showed weak superstructure reflections indicative of chiral K-O-K chains within the channel and a subsequent lowering of the overall crystal symmetry to the triclinic \emph{P$\overline{1}$} (\#2) space group. Smaller crystals grown under conditions similar to those in the previous report often did not display these features and were thus better refined in the hexagonal $P6_3/m$ (\#176) space group with clear diffuse scattering along \textit{b} and more significant in-channel disorder.\cite{eder2023structural} Close analysis of the single crystal diffraction data additionally revealed that KCoTOH crystals natively grow as composites of three coaligned triclinic twins. The twins grow together along the \textit{b}-axis and form an overall hexagonal motif in the \textit{ac}-plane, which is likely responsible for the previous characterization of KCoTOH and related zemannite-type phases in a hexagonal space group. While determination of the atomic positions of hydrogen in the KCoTOH structure is precluded by its small x-ray cross-section, it likely lies entirely within the honeycomb channel as H$_2$O bridges along the K-O-K chains, as has been observed in other zemannite-type structures.\cite{effenberger2023new,johnston2011new} 
Interestingly, the distances between the Co$^{2+}$ cations comprising the pseudo-honeycomb lattice, $d_2$~=~5.5227(5)~--~5.5609(5)~$\Angstrom$, are significantly larger than those observed in any known honeycomb cobaltate and slightly exceed those reported for most triangular lattice systems, with the exception of Ba$_3$CoSb$_2$O$_9$ (Figure \ref{KCoTOH_Main_struc}d, Table \ref{Struc_comp_table}).\cite{istomin2004synthesis} The intra-dimer Co$^{2+}$~-~Co$^{2+}$ distances, $d_1$~=~2.8061(5)~--~2.8355(5)~$\Angstrom$, lie between those observed in the one-dimensional triangular chain system Ca$_3$Co$_2$O$_6$ and triangular dimer K$_2$Co(SeO$_3$)$_2$, while the inter-dimer Co$^{2+}$~-~Co$^{2+}$ distances, $d_3$~=~4.8982(5)~--~4.9303(5)~$\Angstrom$, are shorter than the nearest neighbor distance in any other non-dimerized honeycomb or triangular lattice cobaltate.\cite{miyazaki2005compounds,wildner1992isotypism} Direct Co-O-Co connectivity exists only between cobalts within each structural dimer, with [TeO$_3$]$^{2-}$ groups otherwise separating the dimers along \textit{b}, as well as Co$^{2+}$ within the honeycomb plane, similar to the polyanionic exchange pathways observed in the triangular lattice AFMs Na$_2$BaCo(PO$_4$) and K$_2$Co$_2$(SeO$_3$)$_3$.\cite{zhong2019strong} As in most dimerized systems, the local coordination environment of each Co$^{2+}$ cation in KCoTOH is heavily distorted. Taken together, the unique cobalt-cobalt connectivity in the porous, three-dimensional zemannite-type structure can be considered an effective hybrid between the two more common two-dimensional lattice geometries. 

As has been reported for other zemannite-type materials, the buckled metal tellurate honeycomb framework results in higher thermal stability for KCoTOH than is typically observed in hydrothermally-synthesized phases. Simultaneous thermogravimetric analysis and differential thermal analysis (TGA/DTA) measurements display a minor, stepwise mass decrease up to a congruent melting transition at $T_\text{melt}$~=~595°C, followed by sample decomposition above 800°C (Figure \ref{TGADTA}). Assuming a formula mass of 772.9 g/mol for deuterated KCoTOD, the total mass drop up to the melting transition corresponds to a loss of $\sim$~0.58 D$_2$O molecules, or roughly 23\% of the total D$_2$O content per formula unit, significantly less than would be expected in the case of full dehydration.

\subsection*{Magnetic characterization}

The hybrid triangular dimer/pseudo-honeycomb arrangement of Co$^{2+}$ cations in the KCoTOH structure presents a unique opportunity to study frustrated magnetism at their intersection. While Ising-type magnetic moments are typical in dimerized and triangular lattice cobaltate systems,\cite{zhong2020frustrated2,xu2024frustrated} this is less-commonly observed in antiferromagnetic honeycomb cobaltates.\cite{nair2018short} In the three-dimensional zemannite-type structure, possessing elements of each of these more intensely-studied structure types, magnetic interactions both within and between each [Co$_2$O$_9$] dimer, as well as between Co$^{2+}$ cations in neighboring dimer chains which comprise the pseudo-honeycomb lattice, must all be considered.

\begin{figure}
    \centering
    \includegraphics[width=1.0\linewidth]{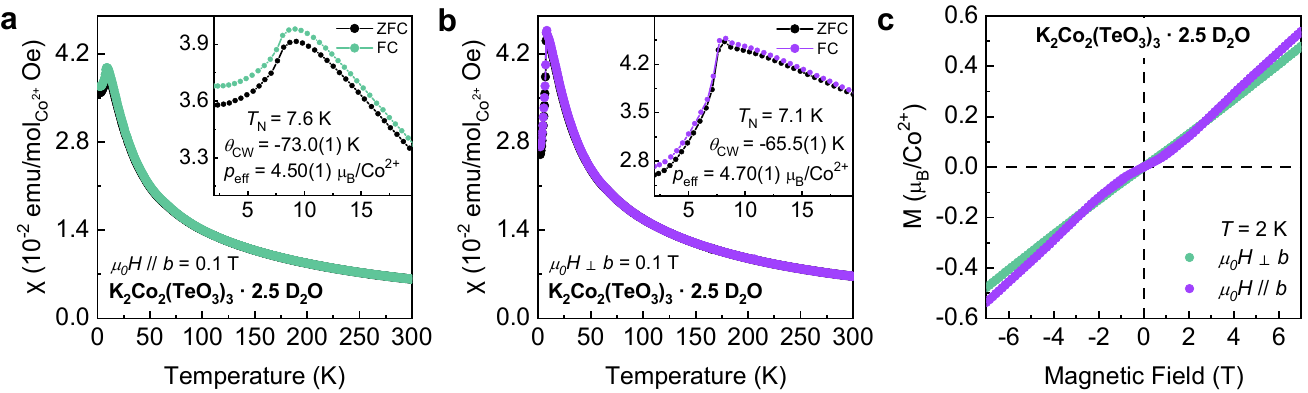}
    \vspace{-10mm}
    \caption{\textbf{Magnetic susceptibility as a function of temperature for representative KCoTOD single crystals} oriented with a) $\mu_\text{o}H~//~b$ and b) $\mu_\text{o}H\perp b$. c) Isothermal, field-dependent magnetization of representative KCoTOD single crystals oriented with $\mu_\text{o}H\perp b$ (green) and $\mu_\text{o}H~//~b$ (purple) at \textit{T}~=~2~K.}
    \label{Mag}
\end{figure}

To explore the magnetic interactions intrinsic to this hybrid structure, the magnetic susceptibility as a function of temperature was measured for coaligned, deuterated KCoTOD crystals. When the field is applied along the \textit{b}-axis, $\mu_\text{o}H~//~b$, a sharp AFM ordering transition is observed at \emph{T}$_\text{N}$~=~7.6~K, with a slight splitting observed between the zero-field-cooled (ZFC) and field-cooled (FC) susceptibility curves, suggestive of the onset of glassy spin dynamics below \textit{T}~=~20~K (Figure \ref{Mag}b). The calculated effective magnetic moment of $p_{\text{eff}}$~=~4.50(1)~$\mu_\text{B}$/Co$^{2+}$ and Curie-Weiss temperature of $\theta_{\text{CW}}$~=~-73.0(1)~K support a \textit{d}$^7$ electron configuration with high spin on each cobalt and net antiferromagnetic exchange along the dimer-chain axis (Table \ref{Oriented_CW}). When the field is applied within the \textit{ac}-plane, $\mu_\text{o}H\perp b$, the AFM ordering transition appears as a sharper cusp at slightly lower temperature, $T_\text{N}$~=~7.1~K, with a greater drop in susceptibility on further cooling (Figure \ref{Mag}a) and negligible splitting between the ZFC and FC curves. However, the calculated effective magnetic moment and Curie-Weiss temperature are quite similar to those observed for $\mu_\text{o}H~//~b$, at $p_{\text{eff}}$~=~4.70(1)~$\mu_\text{B}$/Co$^{2+}$ and $\theta_{\text{CW}}$~=~-65.5(1)~K, respectively (Table \ref{Oriented_CW}). The magnitude of $p_{\text{eff}}$ in both orientations is greater than the $p_{\text{eff}}$~=~3.88~$\mu_\text{B}$/Co$^{2+}$ expected in the spin-only limit and roughly one Bohr magneton smaller than is observed in most honeycomb and triangular lattice cobaltates, suggesting a high-spin configuration with moderate spin-orbit coupling.\cite{zhong2020weak,viciu2007structure} Minimal anisotropy is also observed in the measured magnetization as a function of applied field strength for KCoTOD single crystals at \textit{T}~=~2~K, with the $\mu_\text{o}H~//~b$ dataset being entirely linear, and the $\mu_\text{o}H\perp~b$ data exhibiting only slight curvature between $\mu_\text{o}H$~=~$\pm$~2~T (Figure \ref{Mag}c).

The relative isotropy of the magnetic response of KCoTOH in both field orientations suggests that the magnetic ground state in KCoTOH is driven largely by exchange anisotropy, as opposed to the strong single-ion anisotropy which typically dominates in Co$^{2+}$-based materials. A system with dominant single-ion anisotropy along the dimer chains should result in a more highly anisotropic magnetic response, such as that observed in related K$_2$Co$_2$(SeO$_3$)$_3$.\cite{zhong2020frustrated} This would similarly be the case if magnetic exchange between the dimer chains was negligible - within each high-spin Co$^{2+}$ dimer, both ferromagnetic (FM) direct exchange between each cation and FM superexchange mediated by the bridging O$^{2-}$ \textit{2p}-orbitals [$\angle_{Co-O-Co}$~=~81.9(1)°] between them favor a net parallel alignment of the magnetic moments. However, similar to the extended pathways attributed to meaningful third-nearest neighbor superexchange in honeycomb BaCo$_2$(AsO$_4$)$_2$, significant AFM and FM superexchange likely occurs through the highly-covalent [TeO$_3$]$^{2-}$ groups bridging cobalts within the pseudo-honeycomb plane [$\angle_{Co-Te-Co,1}$ = 158.04(2)°] and between each dimer along \textit{b} [$\angle_{Co-Te-Co, 2}$ = 92.96(2)°], respectively.\cite{kanamori1959superexchange,halloran2023geometrical} With greater orbital overlap along both extended superexchange pathways, the magnetic ground state of KCoTOH is best attributed to the competition between net FM intra- and inter-dimer and AFM intra-honeycomb exchange interactions (occurring along $d_1$, $d_2$, and $d_3$, respectively), resulting in a moderately-frustrated antiferromagnet (with anisotropic frustration parameters \textit{f}~=~9.2 for $\mu_\text{o}H\perp~b$, 9.6 for $\mu_\text{o}H~//~b$).

\subsection*{Specific heat}

\begin{figure}
    \centering
    \includegraphics[width=0.8\linewidth]{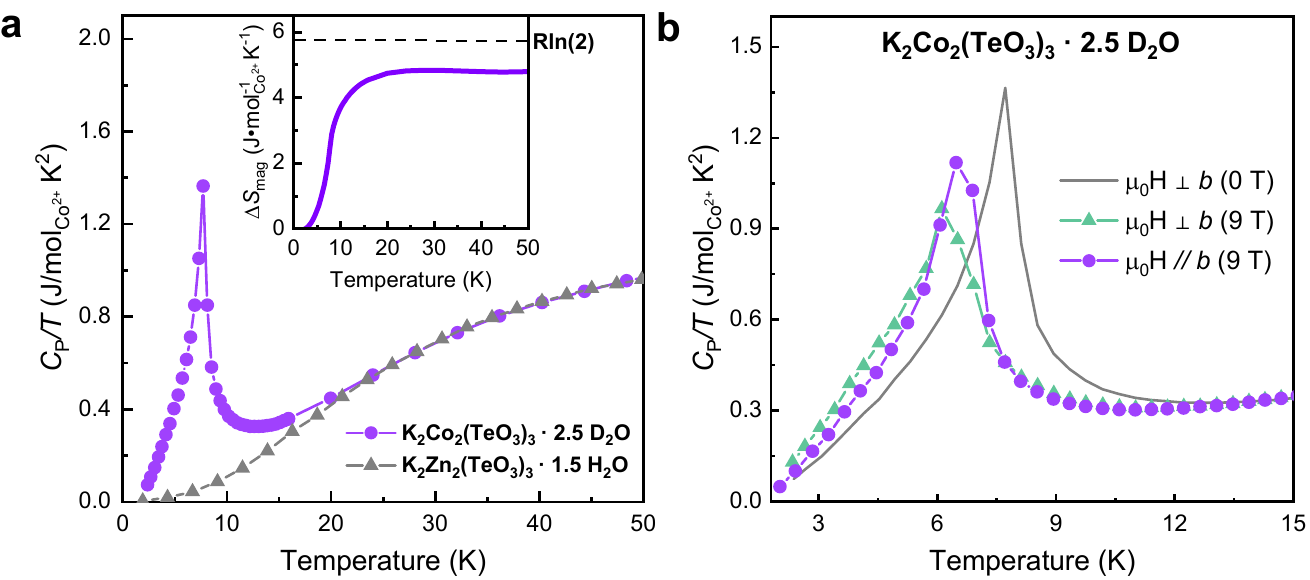}
    \caption{\textbf{Temperature-dependent specific heat of representative KCoTOD single crystals.} a) Representative zero-field specific heat divided by temperature as a function of temperature of KCoTOD (purple) and non-magnetic KZnTOH (gray), as well as magnetic entropy rise for KCoTOD as a function of temperature (inset). b) Specific heat as a function of temperature of KCoTOD single crystals in zero-field (black), as well as in a $\mu_\text{o}H$~=~9~T field applied within the \textit{ac}-plane (purple) and along \textit{b} (green).}
    \label{HC}
\end{figure}

To better understand the low-temperature magnetic behavior of the KCoTOH system, its specific heat was also measured as a function of temperature (Figure \ref{HC}a). When measured in the absence of an applied field, KCoTOD displays a single, sharp feature centered around \textit{T}~=~7.7~K, in agreement with the measured susceptibility data. The application of a $\mu_\text{o}H$~=~9~T field within the \textit{ac}-plane ($\mu_\text{o}H\perp~b$) results in the suppression of this feature to slightly lower temperature (\textit{T}~=~6.1~K) than when the same field is applied along the \textit{b}-axis ($\mu_\text{o}H~//~b$, \textit{T}~=~6.5~K). This weak anisotropy is similar in scale to that observed in the field-dependent magnetization data and is again consistent with dominant AFM interactions within the pseudo-honeycomb plane. Scaled subtraction of the phononic contribution to the measured specific heat, estimated from that of K$_2$Zn$_2$(TeO$_{3}$)$_{3}$~$\cdot$~1.5~(H$_2$O) (KZnTOH), the nonmagnetic analogue to KCoTOD, and subsequent integration of $C_{\text{mag}}$ yields a magnetic entropy rise of $\Delta S_{\mathrm{mag}}$~=~4.83~$J\cdot mol_{Co^{2+}}^{-1}K^{-1}$ up to \textit{T}~=~30~K (Figure \ref{HC}b), with a small amount of entropy missing in the $T~\rightarrow~0$ limit. Above \textit{T}~=~50~K, the two datasets begin to deviate, likely due to the difference in water content between the Co- and Zn-analogues. While lower than expected for either high- or low-spin Co$^{2+}$, where $\Delta S_{\mathrm{mag}}$ = Rln(2S+1) = 11.53 $J\cdot mol^{-1}K^{-1}$ or 5.76 $J\cdot mol^{-1}K^{-1}$ for $S_\text{eff}$~=~3/2 or $S_\text{eff}$~=~1/2 systems, respectively, this is consistent with values reported for other triangular, honeycomb, and triangular dimer lattice cobaltates.\cite{zhong2019strong,zhong2018field,zhong2020frustrated2} 

\subsection*{Neutron diffraction}
Elastic neutron diffraction data collected from a KCoTOH single crystal composite provides further evidence for an ordered magnetic ground state. Due to the morphology and low mass of the as-synthesized samples, the superstructure reflections observed in our SCXRD data could not be reliably resolved in the neutron spectra, resulting in the refinement of the nuclear structure of KCoTOH in the $P6_3/m$ (\#176) space group. As each reflection in the hexagonal structure has more than one possible equivalent in the triclinic notation, two separate notation schemes are required. All Bragg reflections are thus denoted by either (\textit{hkl})$_\text{t}$ or (\textit{hkl})$_\text{h}$, corresponding to the triclinic and hexagonal structures, respectively. 

\begin{figure}
    \centering
    \includegraphics[width=0.8\linewidth]{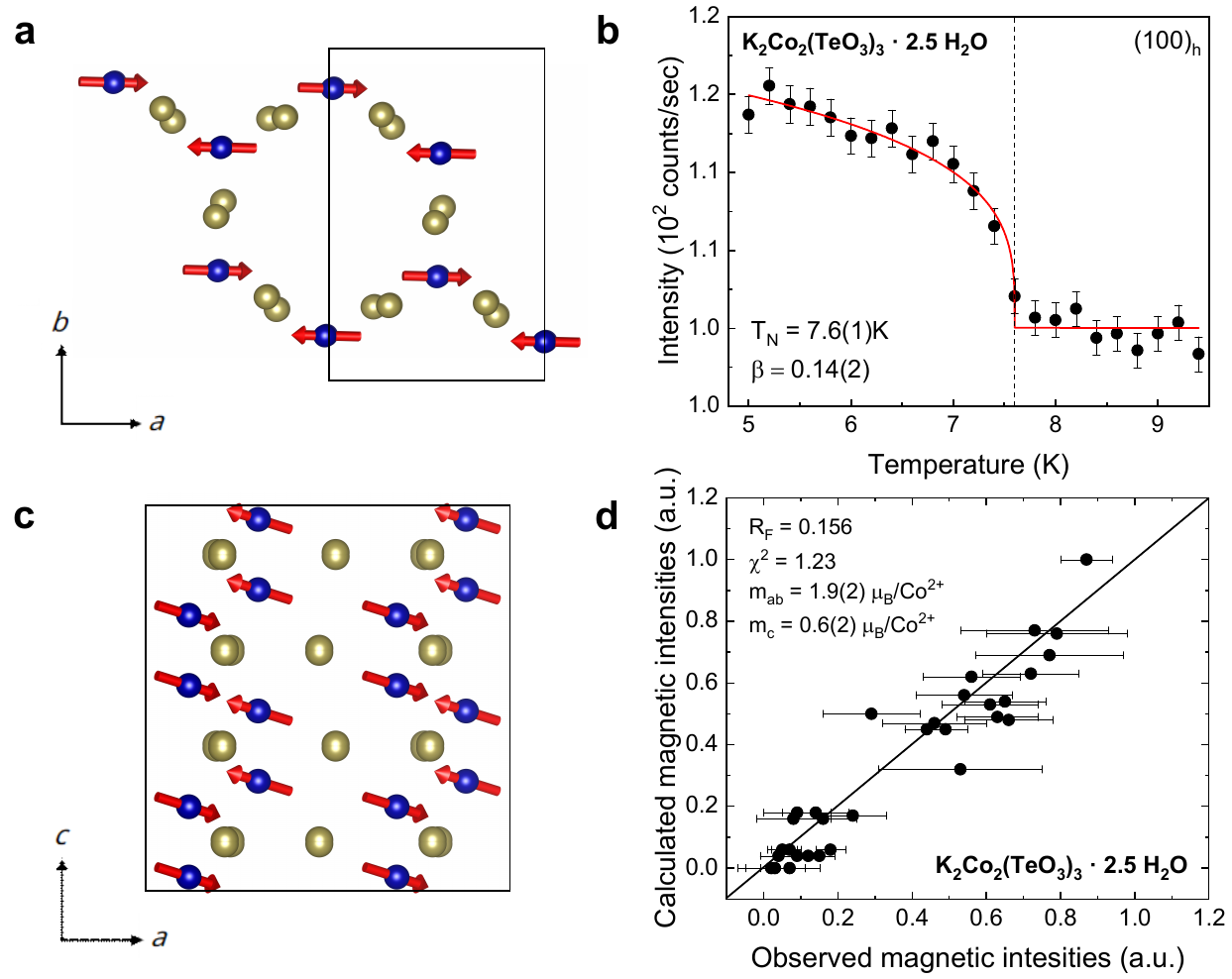}
    \caption{\textbf{Magnetic structure of the ordered ground state in KCoTOH}, shown a) within the pseudo-honeycomb plane and c) along the \textit{c}-axis. Note that the \textit{b}- and \textit{c}-axis labels are inverted, relative to the notation used in Figure 1. b) Magnetic order parameter upon warming of the (1,0,0)$_\text{h}$ reflection as a function of temperature. The red line represents the power-law fit used to calculate $\beta$. d) Fit of the observed vs. calculated intensities of magnetic Bragg reflections for the proposed magnetic ground state.}
    \label{Neutron1}
\end{figure}

Upon cooling through $T_\text{N}$~=~7.6~(1)~K, no additional reflections are observed, while the integrated intensities of most nuclear Bragg reflections increase slightly, indicative of a bulk phase transition to a long-range, AFM ordered state commensurate with our SCXRD-derived structure (Figure \ref{Neutron1}b). For Q~=~0 magnetic order in the $P6_3/m$ space group, there exist 22 possible irreducible representations of the magnetic ground state, 11 of which do not allow for any component of the ordered moment to lie within the \textit{ac}-plane (Table \ref{Irrep_table}). Because neutron scattering is sensitive to the magnetic moments perpendicular to wavevector transfer, these are precluded by the appreciable increase in intensity of the $(006)_h$ Bragg reflection below $T_\text{N}$ (Figure \ref{Neutron2}a). Of the remaining irreps, only 6 represent an AFM-ordered ground state, with refinement in the $P2{_1}/m'$ (\#11.53) and $P\overline{1}'$ (\#2.6) magnetic space groups yielding the best fits to experiment, with $\chi^2$ values of 1.23 and 1.19, respectively. Both support a largely in-plane-ordered moment with an overall AFM arrangement between FM-aligned Co$^{2+}$ dimers, however the former does not allow for the relative isotropy in the observed magnetic response. We therefore conclude that the $P\overline{1}'$ (\#2.6) magnetic space group provides the best overall description of KCoTOH's magnetic ground state.

The critical exponent of the order parameter for this transition is found to be $\beta$~=~0.14(2), consistent with the value expected for a 2D-Ising system ($\beta$~=~$\frac{1}{8}$).\cite{schultz1964two} The majority of the ordered moment, 1.9(2)~$\mu_\text{B}$, lies within the pseudo-honeycomb plane, with the remaining 0.6(2)~$\mu_\text{B}$ canted along the chain axis (Figure \ref{Neutron1}a,c). The net intra-chain FM and inter-chain AFM interactions in this highly three-dimensional structure then give rise to a largely two-dimensional AFM ordered state within the pseudo-honeycomb plane, with only a $\sim$15 degree canting along \textit{b}. While the native twinned morphology of the KCoTOH composites precludes determination of potential in-plane anisotropy, these results are consistent with a $Q=0$ Ne\'el-type AFM ground state, akin to that realized for $J_2<0.2~J_1$ in a $J_1-J_2$ model on the pristine honeycomb lattice with spin-$1/2$ moments \cite{bishop2013valence}. 

\subsection*{muSR characterization}

\begin{figure}[ht]
    \includegraphics[width=\textwidth]{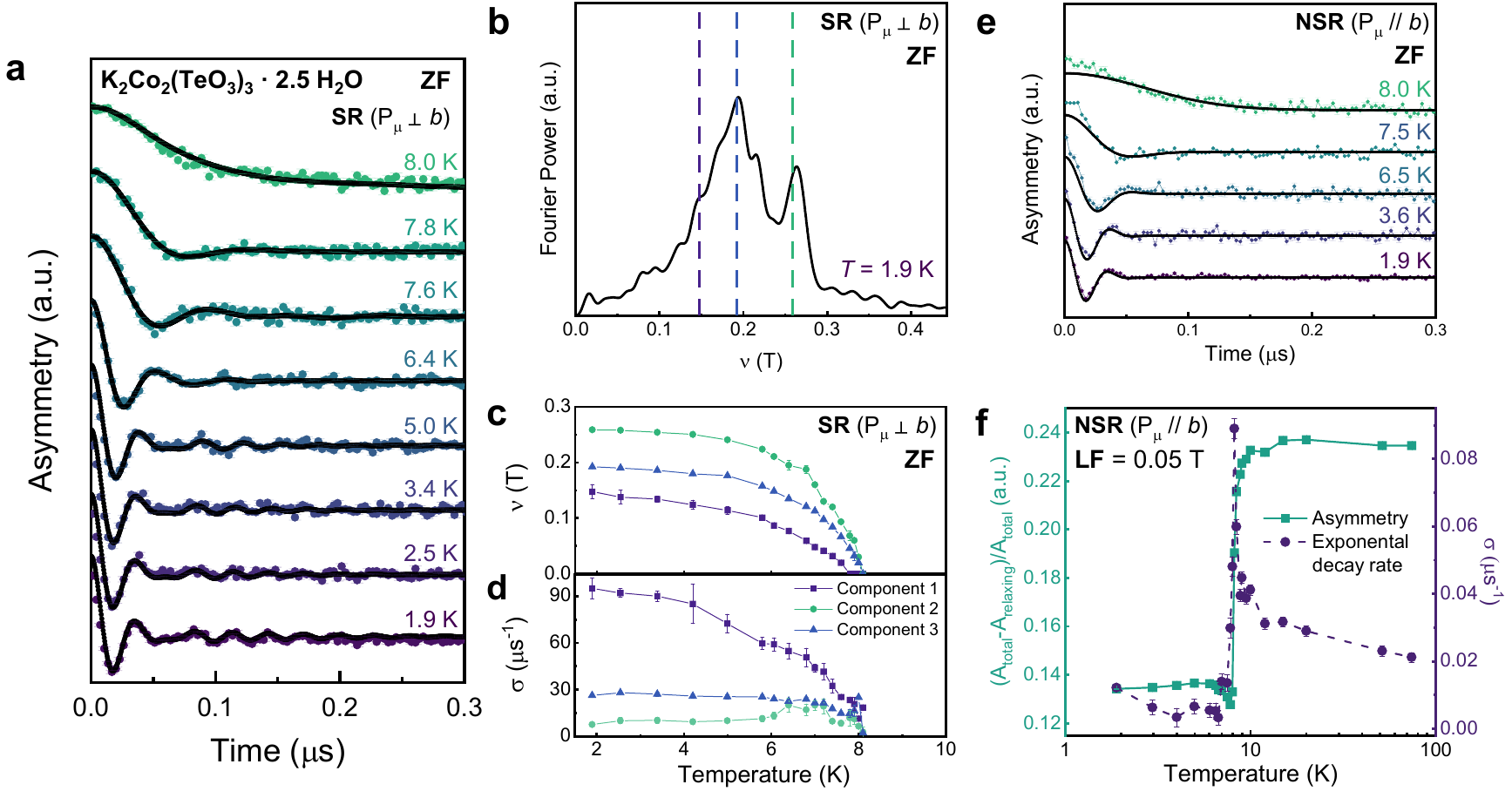}
    \vspace{-12mm}
    \caption{\textbf{Internal magnetic field dynamics for an ensemble of KCoTOH single crystals.} a) Temperature-dependent zero-field (ZF) muon spin relaxation spectra collected from KCoTOH single crystals in the spin-rotated (SR) geometry ($p_{\mu}~\perp~b$). b) Fourier power as a function of frequency for the SR spectra, supporting the presence of at least three distinct magnetic phases at \textit{T}~=~1.9~K, the frequency-dependence and fluctuation rates of which are shown in c) and d), respectively. e) Temperature-dependent zero-field (ZF) muon spin relaxation spectra collected from KCoTOH single crystals in the non-spin-rotated (NSR) geometry ($p_{\mu}~//~b$). f) Fit parameters as a function of temperature derived from longitudinal field (LF) muon spin relaxation spectra collected from KCoTOH single crystals in the NSR geometry, supporting the sharp onset of long-range AFM order below \textit{T}~=~8.0~K.}
    \label{muSR_SR}
\end{figure}

To further probe the magnetic ground state of zemannite-type KCoTOH, zero-field (ZF) muon spin relaxation ($\mu$SR) measurements of coaligned KCoTOD single crystals were performed, both above and below $T_\text{N}$ (Figure \ref{muSR_exp_setup}). ZF spectra collected in the spin-rotated (SR) geometry ($p_{\mu}~\perp~b$) display increasing relaxation and clear oscillations upon cooling through $T_\text{N}$, which signifies the development of static, long-range magnetic order (Figure \ref{muSR_SR}a). The profile of the fast Fourier transform of the \textit{T}~=~1.9~K spectrum clearly shows peaks at at least two distinct frequencies - one sharply centered around $\nu$~=~0.26(1)~T and another broader feature centered around $\nu$~=~0.19(1)~T (Figure \ref{muSR_SR}b). While initial attempts to fit the data to a two-component model yielded general agreement with the very low time ($\geq$~0.05~$\mu$s) decay asymmetry, it could not reproduce the longer-\textit{t} oscillations. The fitting of the ZF muon decay asymmetry was ultimately achieved via the inclusion of a third component as:  
\begin{equation}
    A(t)~=~C_1~+~C_2~+~C_3~+~A_{constant}~~~~~~~~~~~~~~~~C_x~=~A_x\cdot~e^{-\frac{1}{2}(\sigma_x t)^2}\cdot~cos(\gamma_\mu\nu_x t+\varphi_x)
\end{equation}
where three distinct, resolvable relaxing components together account for approximately 79\% of the total asymmetry observed in this orientation and accurately capture the longer-time oscillations. The remaining 21\% of the total asymmetry, $A_{constant}$, is in good agreement with the ~24\% of the ordered moment expected to be perpendicular to the incident muon spin (Figure \ref{muSR_SR},c). The largest of the three components ($C_1,\sim$60\% of relaxing asymmetry) relaxes quickly, as would be expected for a typical ordered AFM (Figure \ref{muSR_SR}d). When considered together with $C_2$ ($\sim$33\% of relaxing asymmetry) and the slowest-relaxing $C_3$ ($\sim$7\% of relaxing asymmetry), these spectra support the coexistence of at least three distinct muon stopping sites in the KCoTOH structure, down to \textit{T}~=~1.9~K (Table \ref{muSR_params_ZF}). The ability to clearly resolve multiple distinct, static components suggests a surprisingly low level of structural disorder and high magnetic homogeneity for an intrinsically-hydrated material grown from solution. 

ZF spectra collected in the non-spin-rotated (NSR) geometry ($p_{\mu}~//~b$) display a similarly sharp change in relaxation rate below $T_\text{N}$, albeit with weaker oscillations at base temperature, due to the smaller component of the ordered moment oriented parallel to that of the incident muon (perpendicular to the pseudo-honeycomb plane) (Figure \ref{muSR_SR}d). As a result, only a single oscillatory component accounting for approximately 46\% of the total asymmetry observed in this orientation could be resolved (Equation 2), with a relaxation rate and average internal field strength roughly in agreement with the average of the three components apparent in the SR orientation (Figure \ref{wTF_NSR_ZF}, Table \ref{muSR_params_NSR}). 
 
\begin{equation}
    A(t)~=~A_o\cdot~e^{(-\lambda t)}\cdot~cos(2\pi\nu t+\varphi)+A_{bg}
\end{equation}

The longitudinal field (LF) dependence of the long-time muon decay asymmetry in this orientation provides evidence for an internal field strength on the order of 0.15~-~0.20~T (Figure \ref{LF_fig}b,c; Table \ref{muSR_params_NSR_LF_B}). The sharp onset of long-range AFM order at $T_\text{N}$~=~7.6~K is evidenced by a distinct slope change in the long-time muon decay asymmetry above and below this temperature in a LF~=~0.49~T field (Figure \ref{muSR_SR}f, Figure \ref{LF_fig}a, Table \ref{muSR_params_NSR_LF_T}). This is also apparent in spectra measured in both the NSR and SR geometries in the presence of a weak transverse field (wTF) (Figure \ref{wTF_NSR_SR_comb}). Above $T_\text{N}$, the full asymmetry oscillates at the frequency of the external field, as expected for a material in its paramagnetic state, while below $T_\text{N}$ the internal field from the sample exceeds the external field, such that the decay asymmetry resembles the zero field measurement. (Table \ref{wTF_fits}). 

\subsection*{Conclusion}

In conclusion, we report the study of novel zemannite-type K$_2$Co$_2$(TeO$_{3}$)$_{3}$~$\cdot$~2.5~H$_2$O (KCoTOH), a hybrid triangular dimer, pseudo-honeycomb cobaltate possessing competing structural and magnetic dimensionalities. Combined single crystal X-ray and neutron diffraction reveals the true triclinic structure of the zemannite-type phase and its native threefold-twinned growth morphology. Despite possessing dimerized chains of Co$^{2+}$ cations, magnetization, specific heat, and muon spin relaxation measurements suggest a 2D-Ising-type AFM magnetic ground state, with the majority of the ordered moment oriented not along the dimer-chain axis, but rather within the pseudo-honeycomb plane. Below $T_\text{N}$, three distinct, homogeneous internal fields are resolved in zero-field $\mu$SR measurements, corresponding to the presence of at least three distinct muon stopping sites in the KCoTOH structure and indicative of unusually low structural disorder for a solution-grown, hydrated material. This work highlights the importance of exploratory synthesis and novel structure types in the development of new frustrated magnets. 

\newpage

\begin{table}
    \centering
    \renewcommand{\arraystretch}{1.2}
    \caption{\textbf{Structural comparison between selected triangular lattice, triangular dimer, and honeycomb lattice cobaltate quantum magnets.} For non-dimerized systems, $d_3$ was taken as the shortest distance between Co$^{2+}$ cations in adjacent layers.} 
    \vspace{1mm}
    \label{Struc_comp_table}
    \renewcommand{\arraystretch}{0.75}
    \begin{tabular}{|c|c|c|c|c|}
    \hline
      \textbf{Phase} & \textbf{Lattice} & $d_1$ & $d_2$ & $d_3$ \\
      & \textbf{Geometry} & ($\Angstrom$) & ($\Angstrom$) & ($\Angstrom$) \\\hline
     Na$_2$Co$_2$TeO$_6$ & Honeycomb\cite{viciu2007structure} & - & 3.0432(1) & 5.6308(8) \\\hline
     Na$_3$Co$_2$SbO$_6$ & Honeycomb\cite{viciu2007structure} & - & 3.0956(1) & 5.637(2) \\\hline
     BaCo$_2$(AsO$_4$)$_2$ & Honeycomb\cite{djordevic2008baco2} & - & 2.8954(6) & 7.994(2)\\\hline
     Na$_2$BaCo(PO$_4$) & Triangular\cite{zhong2019strong} & - & 5.3185(1) & 7.0081(1) \\\hline
     K$_2$Co(SeO$_3$)$_2$ & Triangular\cite{wildner1992isotypism} & - & 5.5160(8) & 6.946(2) \\\hline
     Ba$_3$CoSb$_2$O$_9$ & Triangular\cite{istomin2004synthesis} & - & 5.8532(1) & 7.2227(1) \\\hline 
     Ca$_3$Co$_2$O$_6$ & Triangular chain\cite{miyazaki2005compounds} & 2.5940(1) & 5.5163(2) & 2.5940(1) \\\hline
     K$_2$Co$_2$(SeO$_3$)$_3$ & Triangular dimer\cite{wildner1994structure} & 2.957(2) & 5.486(1) & 6.635(2) \\\hline
     BaCo$_2$(SeO$_3$)$_3$~$\cdot$~3H$_2$O & Zemannite\cite{liu2022magnetism} & 2.883(2) & 4.730(2) & 5.276(1) \\\hline
     K$_2$Co$_2$(TeO$_{3}$)$_{3}$~$\cdot$~2.5~H$_2$O & Zemannite$^*$ & 2.8061(5) -- & 5.5227(5) -- & 4.8982(5) -- \\
     & & 2.8355(5) & 5.5609(5) & 4.9303(5) \\\hline
     \multicolumn{1}{l}{*This work}\\
    \end{tabular}
\end{table}
	


\clearpage 

%
\bibliography{science_template} 

@article{johnston2011new,
  title={New BaM$_2$(SeO$_3$)$_3$~{\textperiodcentered}~n~H$_2$O (M~=~Co,~Ni,~Mn,~Mg; n~$\approx$~3) Zemannite-Type Frameworks: Single-Crystal Structures of BaCo$_2$(SeO$_3$)$_3$~{\textperiodcentered}~3~H$_2$O, BaMn$_2$(SeO$_3$)$_3$~{\textperiodcentered}~3~H$_2$O and BaMg$_2$(SeO$_3$)$_3$~{\textperiodcentered}~3~H$_2$O},
  author={Johnston, Magnus G and Harrison, William TA},
  year={2011},
  journal={European Journal of Inorganic Chemistry},
  volume={2011},
  number={19},
  pages={2967--2974},
  publisher={Wiley Online Library}
}

@article{eder2023structural,
  title={Structural studies on synthetic A$_{2-x}$[M$_2$(TeO$_3$)$_3$]~\textperiodcentered~n~H$_2$O phases (A= Na, K, Rb, Cs; M= Mn, Co, Ni, Cu, Zn) with zemannite-type structures},
  author={Eder, Felix and Marsollier, Alexandre and Weil, Matthias},
  journal={Mineralogy and Petrology},
  volume={117},
  number={2},
  pages={145--163},
  year={2023},
  publisher={Springer}
}

@article{effenberger2023new,
  title={New insights into the crystal chemistry of zemannite: Trigonal rather than hexagonal symmetry due to ordering within the host-guest structure},
  author={Effenberger, Herta S and Ende, Martin and Miletich, Ronald},
  journal={Mineralogy and Petrology},
  volume={117},
  number={2},
  pages={117--131},
  year={2023},
  publisher={Springer}
}

@article{huang2021crystal,
  title={Crystal-to-crystal transformation of a new selenite compound CaNi$_2$(SeO$_3$)$_3$~{\textperiodcentered}~2~H$_2$O induced by dehydration},
  author={Huang, Xing and Zhao, Zhiying and Zhang, Mengsi and He, Zhangzhen},
  journal={CrystEngComm},
  volume={23},
  number={17},
  pages={3126--3132},
  year={2021},
  publisher={Royal Society of Chemistry}
}

@article{liu2022magnetism,
  title={Magnetism and ESR of the $S_\text{eff}$~=~1/2 antiferromagnet BaCo$_2$(SeO$_3$)$_3$~{\textperiodcentered}~3~H$_2$O with dimer-chain structure},
  author={Liu, XC and Ouyang, ZW and Xiao, TT and Cao, JJ and Wang, ZX and Xia, ZC and He, ZZ and Tong, W},
  journal={Physical Review B},
  volume={105},
  number={13},
  pages={134417},
  year={2022},
  publisher={APS}
}

@article{balents2010spin,
  title={Spin liquids in frustrated magnets},
  author={Balents, Leon},
  journal={Nature},
  volume={464},
  number={7286},
  pages={199--208},
  year={2010},
  publisher={Nature Publishing Group UK London}
}

@article{broholm2020quantum,
  title={Quantum spin liquids},
  author={Broholm, C and Cava, RJ and Kivelson, SA and Nocera, DG and Norman, MR and Senthil, T},
  journal={Science},
  volume={367},
  number={6475},
  pages={eaay0668},
  year={2020},
  publisher={American Association for the Advancement of Science}
}

@article{paddison2017continuous,
  title={Continuous excitations of the triangular-lattice quantum spin liquid YbMgGaO$_4$},
  author={Paddison, Joseph AM and Daum, Marcus and Dun, Zhiling and Ehlers, Georg and Liu, Yaohua and Stone, Matthew B and Zhou, Haidong and Mourigal, Martin},
  journal={Nature Physics},
  volume={13},
  number={2},
  pages={117--122},
  year={2017},
  publisher={Nature Publishing Group UK London}
}

@article{zhong2020frustrated,
  title={Frustrated spin-1/2 dimer compound K$_2$Co$_2$(SeO$_3$)$_3$ with easy-axis anisotropy},
  author={Zhong, Ruidan and Guo, Shu and Nguyen, Loi T and Cava, Robert J},
  journal={Physical Review B},
  volume={102},
  number={22},
  pages={224430},
  year={2020},
  publisher={APS}
}

@article{halloran2023geometrical,
  title={Geometrical frustration versus Kitaev interactions in BaCo$_2$(AsO$_4$)$_2$},
  author={Halloran, Thomas and Desrochers, F{\'e}lix and Zhang, Emily Z and Chen, Tong and Chern, Li Ern and Xu, Zhijun and Winn, Barry and Graves-Brook, Melissa and Stone, Matthew B and Kolesnikov, Alexander I and others},
  journal={Proceedings of the National Academy of Sciences},
  volume={120},
  number={2},
  pages={e2215509119},
  year={2023},
  publisher={National Acad Sciences}
}

@article{miao2024persistent,
  title={Persistent spin dynamics in magnetically ordered honeycomb-lattice cobalt oxides},
  author={Miao, Ping and Jin, Xianghong and Yao, Weiliang and Chen, Yue and Koda, Akihiro and Tan, Zhenhong and Xie, Wu and Ji, Wenhai and Kamiyama, Takashi and Li, Yuan},
  journal={Physical Review B},
  volume={109},
  number={13},
  pages={134431},
  year={2024},
  publisher={APS}
}

@article{kitaev2006anyons,
  title={Anyons in an exactly solved model and beyond},
  author={Kitaev, Alexei},
  journal={Annals of Physics},
  volume={321},
  number={1},
  pages={2--111},
  year={2006},
  publisher={Elsevier}
}

@article{maksimov2022ab,
  title={Ab initio guided minimal model for the “Kitaev” material BaCo$_2$(AsO$_4$)$_2$: Importance of direct hopping, third-neighbor exchange, and quantum fluctuations},
  author={Maksimov, Pavel A and Ushakov, Alexey V and Pchelkina, Zlata V and Li, Ying and Winter, Stephen M and Streltsov, Sergey V},
  journal={Physical Review B},
  volume={106},
  number={16},
  pages={165131},
  year={2022},
  publisher={APS}
}

@article{missen2019crystal,
  title={Crystal chemistry of zemannite-type structures: II. Synthetic sodium zemannite},
  author={Missen, Owen P and Mills, Stuart J and Spratt, John},
  journal={European Journal of Mineralogy},
  volume={31},
  number={3},
  pages={529--536},
  year={2019},
  publisher={GeoScienceWorld}
}

@article{halloran2025continuum,
  title={Continuum of magnetic excitations in the Kitaev honeycomb iridate D$_3$LiIr$_2$O$_6$},
  author={Halloran, Thomas and Wang, Yishu and Plumb, KW and Stone, MB and Winn, Barry and Graves-Brook, MK and Rodriguez-Rivera, JA and Qiu, Yiming and Chauhan, Prashant and Knolle, Johannes and others},
  journal={npj Quantum Materials},
  volume={10},
  number={1},
  pages={1--7},
  year={2025},
  publisher={Nature Publishing Group}
}

@article{banerjee2017neutron,
  title={Neutron scattering in the proximate quantum spin liquid $\alpha$-RuCl$_3$},
  author={Banerjee, Arnab and Yan, Jiaqiang and Knolle, Johannes and Bridges, Craig A and Stone, Matthew B and Lumsden, Mark D and Mandrus, David G and Tennant, David A and Moessner, Roderich and Nagler, Stephen E},
  journal={Science},
  volume={356},
  number={6342},
  pages={1055--1059},
  year={2017},
  publisher={American Association for the Advancement of Science}
}

@article{djordevic2008baco2,
  title={BaCo$_2$(AsO$_4$)$_2$},
  author={Dordevi{\'c}, Tamara},
  journal={Structure Reports},
  volume={64},
  number={9},
  pages={i58--i59},
  year={2008},
  publisher={International Union of Crystallography}
}

@article{wildner1994structure,
  title={Structure of K$_2$Co$_2$(SeO$_3$)$_3$},
  author={Wildner, M},
  journal={Crystal Structure Communications},
  volume={50},
  number={3},
  pages={336--338},
  year={1994},
  publisher={International Union of Crystallography}
}

@article{wildner1992isotypism,
  title={Isotypism of a selenite with a carbonate: structure of the buetschliite-type compound K$_2$Co(SeO$_3$)$_2$},
  author={Wildner, M},
  journal={Crystal Structure Communications},
  volume={48},
  number={3},
  pages={410--412},
  year={1992},
  publisher={International Union of Crystallography}
}

@article{zhong2019strong,
  title={Strong quantum fluctuations in a quantum spin liquid candidate with a Co-based triangular lattice},
  author={Zhong, Ruidan and Guo, Shu and Xu, Guangyong and Xu, Zhijun and Cava, Robert J},
  journal={Proceedings of the National Academy of Sciences},
  volume={116},
  number={29},
  pages={14505--14510},
  year={2019},
  publisher={National Academy of Sciences}
}

@article{miyazaki2005compounds,
  title={Compounds and subsolidus phase relations in the CaO--Co$_3$O$_4$--CuO system},
  author={Miyazaki, Yuzuru and Huang, Xiangyang and Kajitani, Tsuyoshi},
  journal={Journal of Solid State Chemistry},
  volume={178},
  number={10},
  pages={2973--2979},
  year={2005},
  publisher={Elsevier}
}

@article{istomin2004synthesis,
  title={Synthesis and characterization of novel 6-H perovskites Ba$_2$Co$_{2- x}$Sb$_{x}$O$_{6-y}$, 0.6~$\leq$~x~$\leq$~0.8 and x~=~1.33 (Ba$_3$CoSb$_2$O$_9$)},
  author={Istomin, S Ya and Koutcenko, VA and Antipov, EV and Lindberg, F and Svensson, G},
  journal={Materials Research Bulletin},
  volume={39},
  number={7-8},
  pages={1013--1022},
  year={2004},
  publisher={Elsevier}
}

@article{viciu2007structure,
  title={Structure and basic magnetic properties of the honeycomb lattice compounds Na$_2$Co$_2$TeO$_6$ and Na$_3$Co$_2$SbO$_6$},
  author={Viciu, L and Huang, Q and Morosan, E and Zandbergen, HW and Greenbaum, NI and McQueen, T and Cava, RJ},
  journal={Journal of Solid State Chemistry},
  volume={180},
  number={3},
  pages={1060--1067},
  year={2007},
  publisher={Elsevier}
}

@article{cao2018demand,
  title={DEMAND, a dimensional extreme magnetic neutron diffractometer at the high flux isotope reactor},
  author={Cao, Huibo and Chakoumakos, Bryan C and Andrews, Katie M and Wu, Yan and Riedel, Richard A and Hodges, Jason and Zhou, Wenduo and Gregory, Ray and Haberl, Bianca and Molaison, Jamie and others},
  journal={Crystals},
  volume={9},
  number={1},
  pages={5},
  year={2018},
  publisher={MDPI}
}

@article{zhong2020frustrated2,
  title={Frustrated magnetism in the layered triangular lattice materials K$_2$Co(SeO$_3$)$_2$ and Rb$_2$Co(SeO$_3$)$_2$},
  author={Zhong, Ruidan and Guo, Shu and Cava, RJ},
  journal={Physical Review Materials},
  volume={4},
  number={8},
  pages={084406},
  year={2020},
  publisher={APS}
}

@article{xu2024frustrated,
  title={Frustrated Magnetism in a Potential Quantum Material Based on Spin-1/2 Co$^{2+}$ Dimers},
  author={Xu, Xianghan and Chen, Tong and Wang, Haozhe and Xie, Weiwei and Broholm, CL and Cava, Robert J},
  journal={Chemistry of Materials},
  volume={36},
  number={9},
  pages={4157--4163},
  year={2024},
  publisher={ACS Publications}
}

@article{zhong2018field,
  title={Field-induced spin-liquid-like state in a magnetic honeycomb lattice},
  author={Zhong, Ruidan and Chung, Mimi and Kong, Tai and Nguyen, Loi T and Lei, Shiming and Cava, Robert Joseph},
  journal={Physical Review B},
  volume={98},
  number={22},
  pages={220407},
  year={2018},
  publisher={APS}
}

@article{chamorro2020chemistry,
  title={Chemistry of quantum spin liquids},
  author={Chamorro, Juan R and McQueen, Tyrel M and Tran, Thao T},
  journal={Chemical Reviews},
  volume={121},
  number={5},
  pages={2898--2934},
  year={2020},
  publisher={ACS Publications}
}

@article{wildner1993zemannite,
  title={Zemannite-type selenites: crystal structures of K$_2$[Co$_2$(SeO$_3$)$_3$]~{\textperiodcentered}~2~H$_2$O and K$_2$[Ni$_2$(SeO$_3$)$_3$]~{\textperiodcentered}~2~H$_2$O},
  author={Wildner, M},
  journal={Mineralogy and Petrology},
  volume={48},
  number={2},
  pages={215--225},
  year={1993},
  publisher={Springer}
}

@article{zhong2020weak,
  title={Weak-field induced nonmagnetic state in a Co-based honeycomb},
  author={Zhong, Ruidan and Gao, Tong and Ong, Nai Phuan and Cava, Robert J},
  journal={Science Advances},
  volume={6},
  number={4},
  pages={eaay6953},
  year={2020},
  publisher={American Association for the Advancement of Science}
}

@article{schultz1964two,
  title={Two-dimensional Ising model as a soluble problem of many fermions},
  author={Schultz, Theodore D and Mattis, Daniel C and Lieb, Elliott H},
  journal={Reviews of Modern Physics},
  volume={36},
  number={3},
  pages={856},
  year={1964},
  publisher={APS}
}

@article{bishop2013valence,
  title={Valence-bond crystalline order in the \textit{S}~=~$1/2$ J$_1$--J$_2$ model on the honeycomb lattice},
  author={Bishop, RF and Li, PHY and Campbell, CE},
  journal={Journal of Physics: Condensed Matter},
  volume={25},
  number={30},
  pages={306002},
  year={2013},
  publisher={IOP Publishing}
}

@article{armbruster2001crystal,
  title={Crystal structures of natural zeolites},
  author={Armbruster, Thomas and Gunter, Mickey E},
  journal={Reviews in Mineralogy and Geochemistry},
  volume={45},
  number={1},
  pages={1--67},
  year={2001},
  publisher={Mineralogical Society of America}
}

@article{furukawa2013chemistry,
  title={The chemistry and applications of metal-organic frameworks},
  author={Furukawa, Hiroyasu and Cordova, Kyle E and O’Keeffe, Michael and Yaghi, Omar M},
  journal={Science},
  volume={341},
  number={6149},
  pages={1230444},
  year={2013},
  publisher={American Association for the Advancement of Science}
}

@article{nair2018short,
  title={Short-range order in the quantum XXZ honeycomb lattice material BaCo$_2$(PO$_4$)$_2$},
  author={Nair, Harikrishnan S and Brown, JM and Coldren, E and Hester, G and Gelfand, MP and Podlesnyak, A and Huang, Q and Ross, KA},
  journal={Physical Review B},
  volume={97},
  number={13},
  pages={134409},
  year={2018},
  publisher={APS}
}

@article{hao2023machine,
  title={Machine-learning-assisted automation of single-crystal neutron diffraction},
  author={Hao, Yiqing and Feng, Erxi and Lu, Dan and Zimmer, Leah and Morgan, Zachary and Chakoumakos, Bryan C. and Zhang, Guannan and Cao, Huibo},
  journal={Applied Crystallography},
  volume={56},
  number={2},
  pages={519--525},
  year={2023},
  publisher={International Union of Crystallography}
}

@inproceedings{rodriguez1990fullprof,
  title={FULLPROF: a program for Rietveld refinement and pattern matching analysis},
  author={Rodriguez-Carvajal, Juan},
  booktitle={Satellite meeting on powder diffraction of the XV congress of the IUCr},
  volume={127},
  pages={127},
  year={1990},
  organization={Toulouse, France}
}

@article{aroyo2006bilbao,
  title={Bilbao Crystallographic Server. II. Representations of crystallographic point groups and space groups},
  author={Aroyo, Mois I and Kirov, Asen and Capillas, Cesar and Perez-Mato, JM and Wondratschek, Hans},
  journal={Foundations of Crystallography},
  volume={62},
  number={2},
  pages={115--128},
  year={2006},
  publisher={International Union of Crystallography}
}

@article{kanamori1959superexchange,
  title={Superexchange interaction and symmetry properties of electron orbitals},
  author={Kanamori, Junjiro},
  journal={Journal of Physics and Chemistry of Solids},
  volume={10},
  number={2-3},
  pages={87--98},
  year={1959},
  publisher={Elsevier}
}
\bibliographystyle{sciencemag}

%
%
%
%
%
%


\section*{Acknowledgments}

\paragraph*{Funding:}
Work at JHU was supported by the Institute for Quantum Matter, an Energy Frontier Research Center funded by the U.S. Department of Energy, Office of Science, Office of Basic Energy Sciences, under Grant DE-SC0019331. The MPMS3 system used for magnetic characterization was funded by the National Science Foundation, Division of Materials Research, Major Research Instrumentation Program, under Award \#1828490. C.L. and T.C. were supported by the U.S. Department of Energy, Office of Science, Basic Energy Sciences under Award No. DE-SC0024469 and by the Gordon and Betty Moore Foundation under GBMF9456. TMM acknowledges support of the David and Lucile Packard Foundation. A portion of this research used resources at the High Flux Isotope Reactor, a DOE Office of Science User Facility operated by the Oak Ridge National Laboratory. The beam time was allocated to HB-3A (DEMAND) on proposal number IPTS-32082.2. Y.H. and H.C. acknowledge support by the U.S. DOE, Office of Science, Office of Basic Energy Sciences, Early Career Research Program Award KC0402020, under Contract DE-AC05-00OR22725. The NIH (1S10OD030352 to M.A.S.) is gratefully acknowledged for financial support. Work at UBC was supported by the Natural Sciences and Engineering Research Council of Canada (NSERC), the Canadian Institute for Advanced Research (CIFAR), the Sloan Research Fellowships program, and the Killam Accelerator Research Fellowship. K.M.K. acknowledges support from the Canada Natural Science and Engineering Research Council under Discovery Grant No. RGPIN/3467-2021.

\paragraph*{Author contributions:}
A.M.F. synthesized and characterized the materials. A.M.F., M.S., Y.H., H.C., and T.C. performed the single crystal X-ray and neutron diffraction experiments, A.M.F., C.L., and T.A.S. performed the magnetization and specific heat measurements under the supervision of N.D. and T.M.M., and A.M.F., C.L., and M.R.R. performed the $\mu$SR experiments under the supervision of K.M.K. and A.M.H. A.M.F., A.M.H.,  and T.M.M. analyzed the data and interpreted the results. A.M.F. wrote the manuscript, with contributions from all authors.
\paragraph*{Competing interests:}
There are no competing interests to declare.
\paragraph*{Data and materials availability:}
All data needed to evaluate the conclusions in the paper is present in the paper and/or the Supplementary Materials and is openly available at the online repository https://doi.org/10.34863/3kpq-r364.


\subsection*{Supplementary materials}
Materials and Methods\\
Figs. S1 to S8\\
Tables S1 to S18\\
References \textit{(36-\arabic{enumiv})}\\ 


\newpage


\renewcommand{\thefigure}{S\arabic{figure}}
\renewcommand{\thetable}{S\arabic{table}}
\renewcommand{\theequation}{S\arabic{equation}}
\renewcommand{\thepage}{S\arabic{page}}
\setcounter{figure}{0}
\setcounter{table}{0}
\setcounter{equation}{0}
\setcounter{page}{1} 


\begin{center}
\section*{Supplementary Materials for\\ \scititle}

	Austin~M.~Ferrenti$^{1,2,4,5\ast}$,
	Maxime~A.~Siegler$^{1}$,
	Yiqing~Hao$^{3}$,
        Chris~Lygouras$^{2}$,\and
        ~Tong~Chen$^{2}$,
        Tiffany~A.~Soetojo$^{2}$,
        Megan~R.~Rutherford$^{4,5}$,
        Kenji~M.~Kojima$^{7}$,\and
        ~Huibo~Cao$^{3}$,
        Natalia~Drichko$^{2}$,
        Alannah~M.~Hallas$^{4,5,6}$,
        Tyrel~M.~McQueen$^{1,2,8}$\and
	\small$^\ast$Corresponding author. Email: austin.ferrenti@ubc.ca\and

\subsubsection*{This PDF file includes:}
Materials and Methods\\
Figures S1 to S8\\
Tables S1 to S18\\

\newpage


\subsection*{Materials and Methods}
\end{center}

Single crystals of K$_2$Co$_2$(TeO$_{3}$)$_{3}$~$\cdot$~2.5~H$_2$O were grown by a hydroflux method in Teflon-lined hydrothermal reaction vessels (Parr Instrument Company, Model \#4749). 13-17 mmol of cobalt pellets (Kurt J. Lesker, 99.95\%) were first added to the empty Teflon liner, followed by 3.99 mmol TeO$_2$ (Acros Organics, 99+\%) and 2.66 mmol CoCO$_3$ $\cdot$ H$_2$O (Strem Chemicals, 99\%-Co). 12 mL of a 5.0 M solution of K$_2$CO$_3$ (Thermo Scientific, 99\%) in deionized H$_2$O was then added to the liner and sparged with argon gas for about 6 minutes to remove dissolved O$_2$. The headspace in the liner was quickly flushed with additional argon gas and the vessel was sealed. The vessel was heated to 200°C over 2 hours, held for 3 days, and cooled to 25°C over 7 days. After cooling, the samples were rinsed with deionized H$_2$O, filtered using a vacuum funnel, and dried for at least 2 hours under dynamic vacuum in a glovebox antechamber. Bright purple needle-like crystals of K$_2$Co$_2$(TeO$_{3}$)$_{3}$~$\cdot$~2.5~H$_2$O with a maximum size of 4 x 0.5 x 0.5 mm were found to co-precipitate with an amorphous CoO$_x$/Co(OH)$_x$ impurity phase and were stored in an Argon-filled glovebox to avoid further hydration.  
Similarly-sized, deuterated single crystals of K$_2$Co$_2$(TeO$_{3}$)$_{3}$~$\cdot$~2.5~D$_2$O (KCoTOD) were grown by the same method described for the hydrogen-containing phase, but with the substitution of D$_2$O (Cambridge Isotope Laboratories, Inc., D, 99.9\%) for deionized H$_2$O in the initial reaction mixture and rinsing steps.

Initial attempts to synthesize K$_2$Co$_2$(TeO$_{3}$)$_{3}$~$\cdot$~2.5~H$_2$O utilizing 70\% reduced TeO$_2$ and CoCO$_3$ $\cdot$ H$_2$O input masses (1.33 mmol and 0.80 mmol, respectively, all else being the same), resulted in the growth of crystals with a maximum size of 1 x 0.1 x 0.1 mm, significantly smaller and thinner than those grown by the method described in the main text. These samples were prone to more significant hydration, suggesting increased disorder in the honeycomb channels. At this scale, further adjustment of the relative TeO$_2$, CoCO$_3$ $\cdot$ H$_2$O, and cobalt metal ratios added to the Teflon liner from those above ultimately resulted in less formation of the title phase.

To approximate the phononic contribution to the measured specific heat of KCoTOD, its nonmagnetic analogue, K$_2$Zn$_2$(TeO$_{3}$)$_{3}$~$\cdot$~1.5~H$_2$O (KZnTOH), was also synthesized by the method described above. 2.65 mmol of ZnO (NOAH Technologies, 99.999\%) and 3.98 mmol TeO$_2$ (Alfa Aesar, 99.995\%) were added to a Teflon liner with 12 mL of a 5.0 M solution of K$_2$CO$_3$ (VWR, 99\%) in deionized H$_2$O. The mixture was then sparged with argon gas for about 6 minutes, the liner headspace was quickly flushed with additional argon gas, and the vessel was sealed. The vessel was heated to 200°C over 2 hours, held for 3 days, and cooled to 50°C over 6 days. After cooling, the samples were rinsed with deionized H$_2$O, filtered using a vacuum funnel, and dried for at least 2 hours under dynamic vacuum in a glovebox antechamber. Clear, needle-shaped crystals of KZnTOH were easily isolated and stored in an Argon-filled glovebox to avoid further hydration.

All single crystal X-ray diffraction (SCXRD) data were measured at 213(2)~K using either a SuperNova diffractometer (equipped with Atlas detector) with Mo K$\alpha$ radiation ($\lambda$~=~0.71073~$\Angstrom$) for the cobalt-containing phases or a XtaLAB Synergy-R (equipped with a rotating-anode X-ray source and HyPix-6000HE detector) with Cu K$\alpha_1$ radiation ($\lambda$~=~1.54056~$\Angstrom$) for the Zn-analogue, both under the program CrysAlisPro (Versions 1.171.41.93a and 1.171.42.49, Rigaku OD, 2020-2022). The same program was used to refine the cell dimensions and for data reduction. The structures were solved with the program SHELXT-2018/2 and refined on \emph{F$^{2}$} with SHELXL-2018/3$.^{44}$ Analytical numeric absorption correction using a multifaceted crystal model was applied using CrysAlisPro. The temperature of the data collection was controlled using either the Cryojet system (manufactured by Oxford Instruments) or the Cryostream 1000 from Oxford Cryosystems. 

A squeeze refinement was performed on the Zn-analogue, K$_x$Zn$_2$(TeO$_3$)$_3$~$\cdot$~y~H$_2$O,  to determine its approximate K (x) and H$_2$O (y) content within the channel framework, suggesting a total of roughly 53 electrons per asymmetric unit. Assuming full K-occupancy in the structure (x~=~2), as in the case of KCoTOH:

\vspace{-8mm}
\begin{equation}
    53~\text{electron count voids}~=~2*19~\text{electrons}/K^+~+~y*10~\text{electrons}/H_2O \Rightarrow y~\simeq~1.5
\end{equation}
\vspace{-8mm}

This corresponds to a formula unit of K$_2$Zn$_2$(TeO$_3$)$_3$~$\cdot$~1.5~(H$_2$O), suggesting either lesser intrinsic water incorporation or K-deficiencies in as-synthesized KZnTOH, relative to KCoTOH. The measured KZnTOH crystal was also found to be pseudo-merohedrally twinned, with the twin relationship (M = -1 0 0 / 0 1 0 / 0 0 -1) corresponding to a twofold axis along the \textit{b} direction, and the BASF scale factor refining to 0.5325(5).

Neutron diffraction measurements were carried out on the HB-3A (DEMAND) diffractometer at the High Flux Isotope Reactor at Oak Ridge National Laboratory.\cite{cao2018demand} Data was collected in four-circle mode on a KCoTOH single crystal composite mounted using an aluminum pin and cooled using a He closed cycle refrigerator, with an incident neutron wavelength of $\lambda$~=~1.533$\Angstrom$. The data was reduced using the instrument's machine learning-assisted peak search algorithm and refinement of the nuclear and magnetic structure was performed using FULLPROF.2k, Version 7.95.\cite{hao2023machine,rodriguez1990fullprof} Symmetry analysis was performed using the representation analysis programs within the Bilbao Crystallographic Server.\cite{aroyo2006bilbao}

Simultaneous thermogravimetric analysis/differential thermal analysis (TGA/DTA) was performed using a TA Instruments Q600 SDT. The samples were loaded into pre-dried alumina pans and heated to 900°C at a rate of 10°C/minute under inert N$_2$ flow (100 mL/minute). Furnace heating was then stopped and the samples cooled naturally under N$_2$ gas flowing at the same rate.

Magnetization data was collected on a Quantum Design Magnetic Property Measurement System (MPMS3). Magnetic susceptibility was approximated as magnetization divided by the applied magnetic field ($\chi\approx M/H$). Heat-capacity data was collected on a Quantum Design Physical Property Measurement System (PPMS) using the semi-adiabatic pulse technique with a 1\% temperature rise and over three time constants. To account for the mass difference between both Co/Zn and D/H and thus more accurately substract the phononic contribution to the measured heat capacity of the Co phase over the relevant temperature range, the temperature scale of the Zn-analogue was scaled by:

\vspace{-4mm}
\begin{equation}
	\frac{\Theta^3_{KZnTOH}}{\Theta^3_{KCoTOD}} = (\frac{\text{Molar mass of KCoTOD}}{\text{Molar mass of KZnTOH}})^{3/2} = (\frac{\text{3091.7 g/mol}}{\text{2999.5 g/mol}})^{3/2} = 0.955.
	\label{HC_scaling}
\end{equation}
\vspace{-4mm}

The measured $C_{\mathrm{p}}/T$ of KZnTOH was additionally scaled to account for a $\sim$2.0\% mass error to ensure convergence between the two curves by \textit{T}~=~30~K. The change in magnetic entropy as a function of temperature, $\Delta S_{\mathrm{mag}}$, was approximated as $\Delta S_{\mathrm{mag}}$~=~$\int C_{\mathrm{p}}/T\,dT$ of the measured $C_{\mathrm{p,mag}}$ for KCoTOD single crystals ($C_{\mathrm{p, total, KCoTOD}}$ - \text{scaled} $C_{\mathrm{p, total, KZnTOH}}$), from \emph{T}~=~0~-~50~K, with the entropy rise from \emph{T}~=~0~-~2.0~K being estimated from linear extrapolation over this range. 

$\mu$SR measurements were performed on the LAMPF spectrometer at the TRIUMF particle accelerator. For this experiment, 300~-~400 KCoTOD single crystals were coaligned along \textit{b} on silver tape for a total sample mass of $\sim$190~mg. Measurements were performed in both the non-spin-rotated (NSR) and spin-rotated (SR) orientations, with the incident muon spin oriented perpendicular and parallel to the \textit{b}-axis, respectively. Temperature-dependent weak transverse field (wTF) $\mu$SR spectra were fit by: 
\begin{equation}
    A(t)~=~C_1~+~C_2~+~A_{bg}~~~~~~~~~~~~~~~~C_x~=~A_x\cdot~e^{-\frac{1}{2}(\sigma_x t)^2}\cdot~cos(\gamma_\mu\nu_x t+\varphi_x)
\end{equation}
Temperature- and field-dependent longitudinal field (LF) $\mu$SR spectra were fit by: 
\begin{equation}
    A(t)~=~A_o\cdot~e^{(-\lambda t)}~+~A_{bg}
\end{equation}
\clearpage


\subsubsection*{Structural characterization}

\begin{table}[ht]
    \caption{\textbf{Single Crystal Structural Refinement Data}}
    \vspace{1mm}
    \label{SC Data}
    \renewcommand{\arraystretch}{0.8}
   \begin{tabular}{|l|l|l|l|}
   \hline
    & \textbf{K$_8$Co$_8$(TeO$_{3}$)$_{12}$} $\cdot$ & \textbf{K$_8$Co$_8$(TeO$_{3}$)$_{12}$} $\cdot$ & \textbf{K$_8$Zn$_8$(TeO$_{3}$)$_{12}$} $\cdot$ \\ 
    & \multicolumn{1}{|c|}{\textbf{10 H$_2$O}} & \multicolumn{1}{|c|}{\textbf{10 D$_2$O}} & \multicolumn{1}{|c|}{\textbf{6 H$_2$O}} \\\hline
    Formula weight (g $mol^{-1}$) & 3071.6 & 3091.7 & 2999.5 \\
    Crystal system & triclinic & triclinic & trigonal\\
    Space group & \emph{P$\overline{1}$} (\#2) & \emph{P$\overline{1}$} (\#2) & \emph{P6$_3$/m} (\#176)\\
    \emph{a}($\Angstrom${}) & 9.3959(2) & 9.3913(2) & 9.43449(8) \\
    \emph{b}($\Angstrom${}) & 15.4699(3) & 15.4694(2) & 9.43449(8) \\
    \emph{c}($\Angstrom${}) & 16.3979(3) & 16.3954(3) & 7.70778(7) \\ 
    $\alpha$(°) & 89.809(2) & 90.010(1) & 90 \\ 
    $\beta$(°) & 89.625(1) & 89.643(1) & 90 \\
    $\gamma$(°) & 89.833(2) & 90.023(1) & 120 \\
    Volume ($\Angstrom${}$^3$) & 2383.43(8) & 2381.87(3) & 594.15(1) \\
    \emph{Z} & 2 & 2 & 2\\
    Radiation & Mo \emph{K}$\alpha$ & Mo \emph{K}$\alpha$ & Cu \emph{K}$\alpha$\\
    & ($\lambda$ = 0.71073 $\Angstrom${}) & ($\lambda$ = 0.71073 $\Angstrom${}) & ($\lambda$ = 1.54178 $\Angstrom${})\\
    Temperature (K) & 213(2) & 213(2) & 213(1)\\
    Reflections collected/unique & 89956/17980 & 93254/17868 & 13303/444\\
    $R_{int}$ & 0.0253 & 0.0283 & 0.0246\\
    Data/parameters & 17980/687 & 17868/688 & 444/29\\
    Goodness-of-fit & 1.153 & 1.113 & 1.196 \\
    $R_1$ [$F^2$ $>$ $2\sigma(F^2)$]$^a$ & 0.0235 & 0.0245 & 0.0236\\
    $R_1$ [all data] & 0.0277 & 0.0283 & 0.0236\\
    $wR(F^2)$$^b$ & 0.0525 & 0.0505 & 0.0562\\
    Largest diff. peak and hole & 1.386 \& -0.898 & 1.474 \& -1.159 & 1.319 \& -1.023 \\ 
    (e $\Angstrom${}$^{-3}$) & & & \\\hline
    \multicolumn{2}{l}{$^a$R(F)=$\Sigma$$|$$|$F$_o$$|$-$|$F$_c$$|$$|$/$\Sigma$$|$$|$F$_o$$|$}\\
    \multicolumn{2}{l}{$^b$R$_\omega$(F$^2_o$)= [$\Sigma\omega$(F$^2_o$-F$^2_c$)$^2$/$\Sigma\omega$(F$^2_o$)$^2$]$^{1/2}$}
    \end{tabular} 
\end{table}

\begin{figure}
    \centering
    \includegraphics[width=1.0\linewidth]{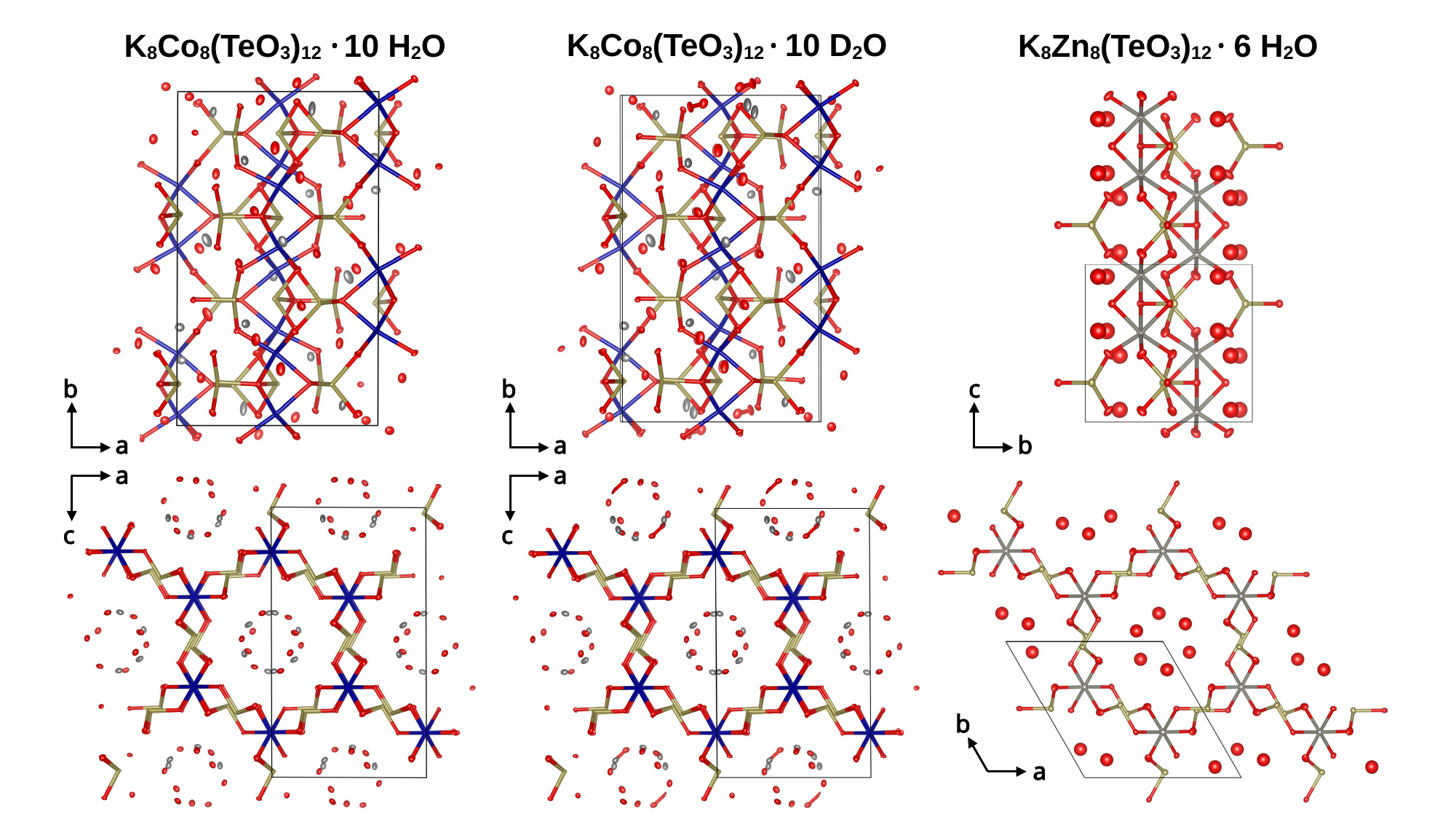}
    \vspace{-10mm}
    \caption{\textbf{Crystal structures of undeuterated (left) and deuterated (center) KCoTOH, as well as undeuterated KZnTOH (right)}, shown both (top) along the chain axis and (bottom) within the pseudo-honeycomb plane. All atoms are shown as anisotropic displacement ellipsoids, the black boxes denote the unit cell of each structure, and K-O bonds have been omitted for clarity.}
    \label{Struc_anis}
\end{figure}

\begin{table}[ht]
    \centering
    \caption{\textbf{K$_2$Co$_2$(TeO$_{3}$)$_{3}$~$\cdot$~2.5~H$_2$O  single crystal structural refinement data - atomic positions and site occupancies}. All $U_{ij}$ values are given in units of $\textrm\AA{^2}$.}
    \vspace{1mm}
    \label{SC2_atomic_pos}
   \renewcommand{\arraystretch}{0.6}
   \begin{tabular}{|c|c|c|c|c|c|}
   \hline
   Atom & x & y & z & Occupancy\\\hline
Te1 & 0.28550(2) & 0.63284(2) & 0.26265(2) & 1 \\
Te2 & 0.28198(2) & 0.12000(2) & 0.26426(2) & 1 \\
Te3 & 1.21633(2) & 0.37564(2) & 0.23692(2) & 1 \\
Te4 & 0.22993(2) & 0.87713(2) & 0.23637(2) & 1 \\
Te5 & 0.78050(2) & 0.37507(2) & 0.24044(2) & 1 \\
Te6 & 0.71594(2) & 0.13409(2) & 0.25697(2) & 1 \\
Te7 & 0.71813(2) & 0.62330(2) & 0.25781(2) & 1 \\
Te8 & -0.20275(2) & 0.87069(2) & 0.24505(2) & 1 \\
Te9 & 0.98640(2) & 0.36843(2) & 0.02204(2) & 1 \\
Te10 & 0.51355(2) & 0.62008(2) & 0.47766(2) & 1 \\
Te11 & -0.00607(2) & 0.87561(2) & 0.02269(2) & 1 \\
Te12 & 0.49641(2) & 0.12564(2) & 0.47766(2) & 1 \\\hline
Co1 & 1.00018(4) & 0.53365(2) & 0.16572(2) & 1 \\
Co2 & 0.49570(4) & 0.46689(2) & 0.33467(2) & 1 \\
Co3 & 0.00946(4) & 0.71685(2) & 0.16534(2) & 1 \\
Co4 & 0.51010(4) & 0.78337(2) & 0.33259(2) & 1 \\
Co5 & 1.00359(4) & 0.21496(2) & 0.16455(2) & 1 \\
Co6 & 0.00519(4) & 1.03357(2) & 0.16673(2) & 1 \\
Co7 & 0.49639(4) & 0.28506(2) & 0.33553(2) & 1 \\
Co8 & 0.50881(4) & -0.03336(2) & 0.33334(2) & 1 \\\hline
K1 & -0.34074(7) & 0.69603(5) & 0.06388(4) & 1 \\
K2 & 0.17638(7) & -0.06099(5) & 0.45557(4) & 1 \\
K3 & 1.02147(8) & 0.19639(5) & 0.39133(4) & 1 \\
K4 & 0.98697(8) & 0.70581(5) & 0.39247(4) & 1 \\
K5 & 0.47805(8) & 0.44965(5) & 0.11114(4) & 1 \\
K6 & 0.33317(7) & 0.79300(5) & -0.04147(4) & 1 \\
K7 & 0.85275(9) & 0.44501(6) & 0.43541(5) & 1 \\
K8 & -0.6661(1) & 1.05187(9) & 0.04163(9) & 0.834(5) \\
K8' & -0.6306(6) & 1.0293(5) & 0.0806(5) & 0.166(5) \\\hline
O1 & 0.9159(2) & 0.1281(1) & 0.2515(1) & 1 \\
O2 & 0.4119(2) & 0.2098(1) & 0.2438(1) & 1 \\
O3 & 0.4018(2) & 0.5313(1) & 0.4335(1) & 1 \\
O4 & 0.0965(2) & 0.7831(1) & 0.0670(1) & 1 \\
O5 & 0.4037(2) & 0.5375(1) & 0.2431(1) & 1 \\
O6 & -0.4032(2) & 0.8767(1) & 0.2483(1) & 1 \\
O7 & 0.6993(2) & 0.7153(1) & 0.3311(1) & 1 \\
O8 & 0.8103(2) & 0.2863(1) & 0.1653(1) & 1 \\
O9 & 0.1071(2) & 0.9692(1) & 0.2652(1) & 1 \\
O10 & 0.4149(2) & 0.7192(1) & 0.2328(1) & 1 \\
O11 & 0.0853(2) & 0.8765(1) & -0.0803(1) & 1 \\\hline
    \end{tabular} 
\end{table}

\begin{table}[ht]
    \centering
    \caption{\textbf{K$_2$Co$_2$(TeO$_{3}$)$_{3}$~$\cdot$~2.5~H$_2$O single crystal structural refinement data - atomic positions and site occupancies}. All $U_{ij}$ values are given in units of $\textrm\AA{^2}$.}
    \vspace{1mm}
    \label{SC2_atomic_pos2}
    \renewcommand{\arraystretch}{0.6}
   \begin{tabular}{|c|c|c|c|c|}
   \hline
   Atom & x & y & z & Occupancy\\\hline
O12 & 1.0782(2) & 0.3747(1) & -0.0801(1) & 1 \\
O13 & 0.1020(2) & 0.7879(1) & 0.2597(1) & 1 \\
O14 & 0.8005(2) & 0.4676(1) & 0.1664(1) & 1 \\
O15 & 0.4006(2) & 0.1241(1) & 0.5788(1) & 1 \\
O16 & 0.4182(2) & 0.6252(1) & 0.5794(1) & 1 \\
O17 & 0.6997(2) & 0.2242(1) & 0.3326(1) & 1 \\
O18 & 0.6891(2) & 0.0448(1) & 0.3337(1) & 1 \\
O19 & 1.0820(2) & 0.4616(1) & 0.2603(1) & 1 \\
O20 & 0.4084(2) & 0.0308(1) & 0.2365(1) & 1 \\
O21 & -0.1767(2) & 0.9623(1) & 0.1712(1) & 1 \\
O22 & 1.3183(2) & 0.3769(1) & 0.3352(1) & 1 \\
O23 & 0.5799(2) & 0.3745(1) & 0.2460(1) & 1 \\
O24 & 1.0973(2) & 0.2799(1) & 0.0673(1) & 1 \\
O25 & 0.1813(2) & 0.6267(1) & 0.1649(1) & 1 \\
O26 & 0.3353(2) & 0.8738(1) & 0.3340(1) & 1 \\
O27 & -0.1810(2) & 0.7834(1) & 0.1679(1) & 1 \\
O28 & 1.0877(2) & 0.4585(1) & 0.0730(1) & 1 \\
O29 & 0.3974(2) & 0.2170(1) & 0.4289(1) & 1 \\
O30 & 0.9186(2) & 0.6224(1) & 0.2527(1) & 1 \\
O31 & 0.4165(2) & 0.7102(1) & 0.4244(1) & 1 \\
O32 & 0.4042(2) & 0.0346(1) & 0.4257(1) & 1 \\
O33 & 0.1808(2) & 0.1233(1) & 0.1655(1) & 1 \\
O34 & 0.3073(2) & 0.4858(2) & -0.0220(1) & 1 \\
O35 & 0.0978(2) & 0.9638(1) & 0.0721(1) & 1 \\
O36 & 0.9421(3) & 0.0155(2) & 0.3937(1) & 1 \\
O37 & 1.0982(2) & 0.2820(1) & 0.2580(1) & 1 \\
O38 & 0.6884(2) & 0.5344(1) & 0.3336(1) & 1 \\
O39 & 0.8801(3) & 0.8576(2) & 0.4606(2) & 1 \\
O40 & 0.1912(2) & -0.2461(2) & 0.5118(2) & 1 \\
O41 & 0.3839(3) & 0.6412(2) & 0.0449(2) & 1 \\
O42 & 1.1516(3) & 0.3322(2) & 0.4653(2) & 1 \\
O43 & -0.5160(3) & 0.8312(2) & 0.0975(2) & 1 \\
O44 & 0.8901(3) & 0.4694(2) & 0.5920(1) & 1 \\
O45 & 0.3914(3) & 0.2574(2) & 0.0895(2) & 1 \\
O46 & -0.4101(3) & 1.0257(2) & 0.1023(2) & 0.834(5) \\
O46' & -0.343(2) & 1.032(1) & 0.067(1) & 0.166(5) \\\hline
    \end{tabular} 
\end{table}

\begin{table}[ht]
    \centering
    \caption{\textbf{K$_2$Co$_2$(TeO$_{3}$)$_{3}$~$\cdot$~2.5~H$_2$O  single crystal structural refinement data - anisotropic displacement parameters}. All $U_{ij}$ values are given in units of $\textrm\AA{^2}$.}
    \vspace{1mm}
    \label{SC2_U_1}
    \renewcommand{\arraystretch}{0.6}
   \begin{tabular}{|c|c|c|c|c|c|c|}
   \hline
   Atom & $U_{11}$ & $U_{22}$ & $U_{33}$ & $U_{12}$ & $U_{13}$ & $U_{23}$\\\hline
Te1 & 0.00720(7) & 0.00738(7) & 0.00725(7) & 0.00036(6) & -0.00094(5) & -0.00016(6) \\
Te2 & 0.00668(7) & 0.00686(7) & 0.00804(7) & 0.00004(6) & -0.00097(5) & -0.00012(5) \\
Te3 & 0.00680(7) & 0.00703(7) & 0.00644(7) & -0.00008(6) & -0.00029(5) & -0.00036(5) \\
Te4 & 0.00673(7) & 0.00638(7) & 0.00736(7) & -0.00051(6) & -0.00074(5) & 0.00003(5) \\
Te5 & 0.00654(7) & 0.00602(7) & 0.00627(7) & 0.00003(5) & 0.00016(5) & 0.00005(5) \\
Te6 & 0.00711(7) & 0.00752(7) & 0.00675(7) & -0.00079(6) & 0.00007(5) & 0.00031(6) \\
Te7 & 0.00655(7) & 0.00673(7) & 0.00682(7) & 0.00004(6) & 0.00023(5) & -0.00007(5) \\
Te8 & 0.00663(7) & 0.00696(7) & 0.00722(7) & -0.00113(6) & -0.00034(5) & -0.00005(5) \\
Te9 & 0.00684(7) & 0.00788(7) & 0.00603(7) & 0.00009(6) & 0.00017(5) & 0.00019(6) \\
Te10 & 0.00712(7) & 0.00729(7) & 0.00609(7) & -0.00023(6) & -0.00023(5) & -0.00021(6) \\
Te11 & 0.00892(7) & 0.00677(7) & 0.00651(7) & -0.00033(6) & 0.00058(6) & -0.00019(6) \\
Te12 & 0.00793(7) & 0.00701(7) & 0.00655(7) & -0.00025(6) & -0.00018(5) & -0.00065(6) \\\hline
Co1 & 0.0077(2) & 0.0053(2) & 0.0061(2) & -0.0001(1) & -0.0001(1) & 0.0002(1) \\
Co2 & 0.0077(2) & 0.0052(2) & 0.0069(2) & -0.0001(1) & -0.0004(1) & 0.0000(1) \\
Co3 & 0.0071(2) & 0.0054(2) & 0.0067(2) & -0.0004(1) & -0.0002(1) & 0.0002(1) \\
Co4 & 0.0074(2) & 0.0052(2) & 0.0074(2) & 0.0001(1) & 0.0000(1) & -0.0001(1) \\
Co5 & 0.0080(2) & 0.0052(2) & 0.0066(2) & 0.0000(1) & -0.0001(1) & -0.0005(1) \\
Co6 & 0.0073(2) & 0.0052(2) & 0.0074(2) & -0.0003(1) & -0.0004(1) & -0.0003(1) \\
Co7 & 0.0078(2) & 0.0052(2) & 0.0066(2) & -0.0003(1) & -0.0002(1) & -0.0003(1) \\
Co8 & 0.0073(2) & 0.0051(2) & 0.0077(2) & -0.0004(1) & -0.0001(1) & 0.0001(1) \\\hline
K1 & 0.0161(3) & 0.0177(3) & 0.0173(3) & -0.0013(2) & -0.0057(2) & 0.0011(2) \\
K2 & 0.0125(3) & 0.0250(4) & 0.0192(3) & 0.0017(3) & 0.0024(2) & 0.0030(2) \\
K3 & 0.0241(3) & 0.0179(3) & 0.0122(3) & -0.0011(2) & 0.0035(2) & -0.0009(3) \\
K4 & 0.0218(3) & 0.0176(3) & 0.0124(3) & -0.0008(2) & -0.0020(2) & -0.0008(2) \\
K5 & 0.0193(3) & 0.0305(4) & 0.0125(3) & 0.0043(3) & 0.0016(2) & -0.0035(3) \\
K6 & 0.0130(3) & 0.0234(3) & 0.0203(3) & -0.0020(3) & -0.0022(2) & -0.0001(2) \\
K7 & 0.0238(4) & 0.0434(5) & 0.0170(3) & 0.0037(3) & -0.0079(3) & -0.0107(3) \\
K8 & 0.0126(4) & 0.0487(7) & 0.0243(7) & -0.0151(5) & 0.0005(4) & 0.0030(4) \\
K8' & 0.018(2) & 0.054(4) & 0.026(3) & -0.011(3) & 0.005(2) & 0.003(2) \\\hline
O1 & 0.0081(8) & 0.0089(9) & 0.0093(9) & -0.0004(7) & 0.0012(7) & -0.0006(7) \\
O2 & 0.0139(9) & 0.0106(9) & 0.0100(9) & -0.0015(7) & 0.0003(7) & -0.0064(7) \\
O3 & 0.0112(9) & 0.0115(9) & 0.0100(9) & -0.0048(7) & -0.0008(7) & -0.0012(7) \\
O4 & 0.0129(9) & 0.0105(9) & 0.0124(9) & 0.0052(7) & 0.0011(7) & 0.0020(7) \\
O5 & 0.016(1) & 0.0106(9) & 0.0113(9) & 0.0012(7) & -0.0010(7) & 0.0051(8) \\
O6 & 0.0063(8) & 0.0090(9) & 0.0097(9) & -0.0010(7) & 0.0018(7) & 0.0004(7) \\
O7 & 0.0119(9) & 0.0113(9) & 0.0119(9) & -0.0046(7) & -0.0012(7) & 0.0042(7) \\
O8 & 0.0117(9) & 0.0114(9) & 0.0106(9) & -0.0038(7) & 0.0001(7) & 0.0042(7) \\
O9 & 0.0118(9) & 0.0096(9) & 0.0129(9) & -0.0004(7) & -0.0010(7) & 0.0042(7) \\
O10 & 0.0134(9) & 0.0102(9) & 0.0126(9) & -0.0005(7) & -0.0017(7) & -0.0046(7) \\
O11 & 0.0114(9) & 0.0082(9) & 0.0067(8) & 0.0013(7) & 0.0011(7) & 0.0003(7) \\\hline
    \end{tabular} 
\end{table}

\begin{table}[ht]
    \centering
    \caption{\textbf{K$_2$Co$_2$(TeO$_{3}$)$_{3}$~$\cdot$~2.5~H$_2$O  single crystal structural refinement data - anisotropic displacement parameters}. All $U_{ij}$ values are given in units of $\textrm\AA{^2}$.}
    \vspace{1mm}
    \label{SC2_U_2}
    \renewcommand{\arraystretch}{0.6}
   \begin{tabular}{|c|c|c|c|c|c|c|}
   \hline
   Atom & $U_{11}$ & $U_{22}$ & $U_{33}$ & $U_{12}$ & $U_{13}$ & $U_{23}$\\\hline
O12 & 0.0107(8) & 0.0068(8) & 0.0051(8) & 0.0000(6) & 0.0008(6) & 0.0005(7)  \\
O13 & 0.0134(9) & 0.0092(9) & 0.0105(9) & -0.0011(7) & -0.0005(7) & -0.0049(7) \\
O14 & 0.0128(9) & 0.0098(9) & 0.0108(9) & 0.0037(7) & -0.0007(7) & -0.0027(7) \\
O15 & 0.0107(9) & 0.0088(9) & 0.0074(8) & 0.0007(7) & 0.0008(7) & 0.0004(7) \\
O16 & 0.0113(9) & 0.0085(9) & 0.0056(8) & -0.0006(7) & 0.0006(7) & -0.0009(7) \\
O17 & 0.0117(9) & 0.0106(9) & 0.0121(9) & -0.0045(7) & -0.0008(7) & 0.0029(7) \\
O18 & 0.0115(9) & 0.012(1) & 0.0119(9) & 0.0039(7) & -0.0001(7) & -0.0037(7) \\
O19 & 0.0143(9) & 0.011(1) & 0.0101(9) & 0.0004(7) & -0.0003(7) & 0.0044(8) \\
O20 & 0.0114(9) & 0.0102(9) & 0.014(1) & 0.0015(7) & -0.0005(7) & 0.0037(7) \\
O21 & 0.0113(9) & 0.013(1) & 0.015(1) & 0.0054(8) & -0.0006(8) & -0.0047(8) \\
O22 & 0.0083(8) & 0.0084(9) & 0.0078(8) & 0.0005(7) & -0.0020(7) & 0.0007(7) \\
O23 & 0.0070(8) & 0.0084(9) & 0.0088(8) & 0.0010(7) & 0.0003(7) & -0.0008(7) \\
O24 & 0.0115(9) & 0.0096(9) & 0.0122(9) & 0.0044(7) & -0.0008(7) & 0.0013(7) \\
O25 & 0.0083(8) & 0.0083(9) & 0.0084(8) & 0.0002(7) & -0.0015(7) & 0.0002(7) \\
O26 & 0.0083(8) & 0.0082(9) & 0.0092(8) & -0.0008(7) & -0.0018(7) & 0.0005(7) \\
O27 & 0.0106(9) & 0.0112(9) & 0.0123(9) & -0.0056(7) & -0.0005(7) & 0.0033(7) \\
O28 & 0.0126(9) & 0.0113(9) & 0.0113(9) & -0.0055(7) & 0.0008(7) & -0.0017(7) \\
O29 & 0.015(1) & 0.012(1) & 0.0128(9) & 0.0054(8) & 0.0008(8) & 0.0010(8) \\
O30 & 0.0068(8) & 0.0087(9) & 0.0085(8) & -0.0006(7) & 0.0013(6) & -0.0003(7) \\
O31 & 0.0114(9) & 0.0111(9) & 0.0130(9) & 0.0054(7) & 0.0005(7) & 0.0024(7) \\
O32 & 0.0115(9) & 0.0107(9) & 0.015(1) & -0.0061(8) & 0.0009(7) & -0.0019(7) \\
O33 & 0.0086(8) & 0.0083(9) & 0.0087(8) & -0.0011(7) & -0.0016(7) & 0.0007(7) \\
O34 & 0.0108(9) & 0.020(1) & 0.019(1) & -0.0012(9) & -0.0014(8) & -0.0036(8) \\
O35 & 0.0140(9) & 0.012(1) & 0.014(1) & -0.0073(8) & 0.0011(8) & -0.0017(8) \\
O36 & 0.020(1) & 0.019(1) & 0.013(1) & 0.0024(8) & -0.0034(8) & 0.0004(9) \\
O37 & 0.0147(9) & 0.0105(9) & 0.0119(9) & -0.0015(7) & -0.0030(7) & -0.0050(8) \\
O38 & 0.0134(9) & 0.012(1) & 0.0107(9) & 0.0046(7) & -0.0008(7) & -0.0047(8) \\
O39 & 0.015(1) & 0.025(1) & 0.021(1) & 0.005(1) & -0.0036(9) & 0.0024(9) \\
O40 & 0.013(1) & 0.032(1) & 0.018(1) & -0.004(1) & -0.0011(8) & 0.0019(9) \\
O41 & 0.017(1) & 0.027(1) & 0.020(1) & -0.003(1) & 0.0037(9) & -0.006(1) \\
O42 & 0.024(1) & 0.044(2) & 0.017(1) & 0.004(1) & 0.002(1) & -0.016(1) \\
O43 & 0.018(1) & 0.037(2) & 0.019(1) & -0.013(1) & -0.0014(9) & 0.002(1) \\
O44 & 0.020(1) & 0.028(1) & 0.015(1) & -0.0029(9) & 0.0031(9) & -0.003(1) \\
O45 & 0.018(1) & 0.040(2) & 0.016(1) & 0.006(1) & 0.0047(9) & 0.001(1) \\
O46 & 0.018(1) &  0.030(2) & 0.014(1) & -0.003(1) & -0.001(1) & 0.006(1) \\
O46' & 0.015(6) & 0.031(7) & 0.027(6) & -0.002(6) & -0.002(5) & 0.002(5) \\\hline
    \end{tabular} 
\end{table}

\begin{table}[ht]
    \centering
    \caption{\textbf{K$_2$Co$_2$(TeO$_{3}$)$_{3}$~$\cdot$~2.5~D$_2$O  single crystal structural refinement data - atomic positions and site occupancies}. All $U_{ij}$ values are given in units of $\textrm\AA{^2}$.}
    \vspace{1mm}
    \label{SC_atomic_pos}
   \renewcommand{\arraystretch}{0.6}
   \begin{tabular}{|c|c|c|c|c|c|}
   \hline
   Atom & x & y & z & Occupancy\\\hline
   Te1 & 0.28522(3) & 0.63261(2) &0.26275(2) & 1 \\
Te2 & 0.28180(2) & 0.12012(2) &0.26430(2) & 1 \\
Te3 & 1.21636(2) &0.37562(2) &0.23682(2) & 1 \\
Te4 & 0.22967(2) &0.87703(2) &0.23628(2) & 1 \\
Te5 & 0.78062(3) &0.37499(2) &0.24042(2) & 1 \\
Te6 & 0.71568(3) &0.13399(2) &0.25688(2) & 1 \\
Te7 & 0.71805(2) &0.62343(2) &0.25786(2) & 1 \\
Te8 & -0.20314(3) &0.87052(2) &0.24504(2) & 1 \\
Te9 & 0.98671(3) &0.36855(2) &0.02202(2) & 1 \\
Te10 & 0.51336(3) &0.62013(2) &0.47771(2) & 1 \\
Te11 & -0.00628(2) &0.87561(2) &0.02262(2) & 1 \\
Te12 & 0.49663(3) &0.12560(2) &0.47772(2) & 1 \\\hline
Co1 & 1.00007(7) &0.53361(4) &0.16581(4) & 1 \\
Co2 & 0.49588(7) &0.46687(4) &0.33462(4) & 1 \\
Co3 & 0.00927(7) &0.71679(4) &0.16538(4) & 1 \\
Co4 & 0.50985(7) &0.78342(4) &0.33265(4) & 1 \\
Co5 & 1.00346(7) &0.21489(4) &0.16458(4) & 1 \\
Co6 & 0.00491(7) &1.03352(4) &0.16678(4) & 1 \\
Co7 & 0.49636(7) &0.28510(4) &0.33540(4) & 1 \\
Co8 & 0.50857(7) &-0.03335(4) &0.33325(4) & 1 \\\hline
K1 & -0.3410(1) &0.69613(7) &0.06390(7) & 1 \\
K2 & 0.1761(1) &-0.06059(7) &0.45546(7) & 1 \\
K3 & 1.0215(1) &0.19638(7) &0.39135(7) & 1 \\
K4 & 0.9870(1) &0.70576(7) &0.39259(7) & 1 \\
K5 & 0.4784(1) &0.44982(8) &0.11112(7) & 1 \\
K6 & 0.3331(1) &0.79281(7) &-0.04125(7) & 1 \\
K7 & 0.8529(1) &0.44494(9) &0.43528(8) & 1 \\
K8 & -0.6660(2) &1.0515(1) &0.0421(1) & 0.821(7) \\
K8' & -0.6304(9) &1.0294(6) &0.0794(7) & 0.179(7) \\\hline  
O1 & 0.9162(3) &0.1275(2) &0.2514(2) & 1 \\
O2 & 0.4119(4) &0.2098(2) &0.2441(2) & 1 \\
O3 & 0.4019(4) &0.5313(2) &0.4337(2) & 1 \\
O4 & 0.0960(4) &0.7837(2) &0.0670(3) & 1 \\
O5 & 0.4026(4) &0.5371(2) &0.2430(2) & 1 \\
O6 & -0.4042(3)& 0.8776(2) &0.2480(2) & 1 \\ 
O7 & 0.6991(4) &0.7154(2) &0.3315(2) & 1 \\
O8 & 0.8105(4) &0.2862(2) &0.1647(2) & 1 \\
O9 & 0.1074(4) &0.9689(2) &0.2650(2) & 1 \\
O10 & 0.4145(4) &0.7188(2) &0.2331(2) & 1 \\
O11 & 0.0857(3) &0.8761(2) &-0.0797(2) & 1 \\\hline
    \end{tabular} 
\end{table}

\begin{table}[ht]
    \centering
    \caption{\textbf{K$_2$Co$_2$(TeO$_{3}$)$_{3}$~$\cdot$~2.5~D$_2$O single crystal structural refinement data - atomic positions and site occupancies}. All $U_{ij}$ values are given in units of $\textrm\AA{^2}$.}
    \vspace{1mm}
    \label{SC_atomic_pos2}
    \renewcommand{\arraystretch}{0.6}
   \begin{tabular}{|c|c|c|c|c|c|}
   \hline
   Atom & x & y & z & Occupancy\\\hline
O12 & 1.0785(3) &0.3738(2) &-0.0803(2) & 1 \\
O13 & 0.1012(4) &0.7880(2) &0.2595(2) & 1 \\
O14 & 0.8004(4) &0.4672(2) &0.1668(2) & 1 \\
O15 & 0.4008(3) &0.1242(2) &0.5784(2) & 1 \\
O16 & 0.4186(3) &0.6257(2) &0.5796(2) & 1 \\
O17 & 0.6995(4) &0.2243(2) &0.3330(2) & 1 \\
O18 & 0.6891(4) &0.0447(2) &0.3331(2) & 1 \\
O19 & 1.0830(4) &0.4616(2) &0.2608(2) & 1 \\
O20 & 0.4079(4) &0.0309(2) &0.2364(2) & 1 \\
O21 & -0.1779(4) &0.9623(2) &0.1718(3) & 1 \\
O22 & 1.3187(3) &0.3768(2) &0.3353(2) & 1 \\
O23 & 0.5805(3) &0.3745(2) &0.2459(2) & 1 \\
O24 & 1.0966(4) &0.2802(2) &0.0675(2) & 1 \\
O25 & 0.1810(3) &0.6265(2) &0.1648(2) & 1 \\
O26 & 0.3354(3) &0.8739(2) &0.3337(2) & 1 \\
O27 & -0.1808(4) &0.7834(2) &0.1676(2) & 1 \\
O28 & 1.0883(4) &0.4583(2) &0.0731(2) & 1 \\
O29 & 0.3978(4) &0.2163(2) &0.4287(3) & 1 \\
O30 & 0.9179(3) &0.6226(2) &0.2524(2) & 1 \\
O31 & 0.4166(4) &0.7101(2) &0.4246(2) & 1 \\
O32 & 0.4035(4) &0.0341(2) &0.4260(2) & 1 \\
O33 & 0.1804(3) &0.1231(2) &0.1657(2) & 1 \\
O34 & 0.3073(4) &0.4860(2) &-0.0217(2) & 1 \\
O35 & 0.0979(4) &0.9643(2) &0.0720(3) & 1 \\
O36 & 0.9419(4) &0.0151(2) &0.3938(2) & 1 \\
O37 & 1.0977(4) &0.2823(2) &0.2581(2) & 1 \\
O38 & 0.6888(4) &0.5351(2) &0.3333(2) & 1 \\
O39 & 0.8793(3) &0.8574(2) &0.4606(2) & 1 \\
O40 & 0.1918(4) &-0.2464(2) &0.5127(2) & 1 \\
O41 & 0.3837(4) &0.6413(2) &0.0450(2) & 1 \\
O42 & 1.1515(4) &0.3322(3) &0.4655(2) & 1 \\
O43 & -0.5160(4) &0.8311(3) &0.0977(2) & 1 \\
O44 & 0.8896(4) &0.4694(3) &0.5921(2) & 1 \\
O45 & 0.3917(4) &0.2570(3) &0.0901(2) & 1 \\
O46 & -0.4111(5) &1.0249(3) &0.1024(3) & 0.821(7) \\
O46' & -0.344(2) &1.036(1) &0.067(1) & 0.179(7) \\\hline   
    \end{tabular} 
\end{table}

\begin{table}[ht]
    \centering
    \caption{\textbf{K$_2$Co$_2$(TeO$_{3}$)$_{3}$~$\cdot$~2.5~D$_2$O  single crystal structural refinement data - anisotropic displacement parameters}. All $U_{ij}$ values are given in units of $\textrm\AA{^2}$.}
    \vspace{1mm}
    \label{SC_U_1}
    \renewcommand{\arraystretch}{0.6}
   \begin{tabular}{|c|c|c|c|c|c|c|}
   \hline
   Atom & $U_{11}$ & $U_{22}$ & $U_{33}$ & $U_{12}$ & $U_{13}$ & $U_{23}$\\\hline
Te1 & 0.0063(1) & 0.0067(1) & 0.0072(1) & 0.0004(1) & -0.00116(8) & 0.0004(1) \\
Te2 & 0.0055(1) & 0.0065(1) & 0.0079(1) & 0.0001(1) & -0.00119(8) & -0.0001(1) \\
Te3 & 0.0059(1) & 0.0060(1) & 0.0066(1) & 0.0001(1) & -0.00045(8) & -0.0002(1) \\
Te4 & 0.0057(1) & 0.0063(1) & 0.0074(1) & -0.0007(1) & -0.00082(8) & 0.0000(1)\\
Te5 & 0.0058(1) & 0.0052(1) & 0.0061(1) & -0.0003(1) & -0.00010(8) & -0.0001(1)\\
Te6 & 0.0061(1) & 0.0066(1) & 0.0067(1) & -0.0007(1) & 0.00006(8) & 0.0007(1)\\
Te7 & 0.0055(1) & 0.0057(1) & 0.0067(1) & -0.0003(1) & -0.00004(8) & -0.0002(1)\\
Te8 & 0.0057(1) & 0.0066(1) & 0.0074(1) & -0.0014(1) & -0.00018(8) & 0.0003(1)\\
Te9 & 0.0061(1) & 0.0076(1) & 0.0057(1) & 0.0002(1) & 0.00010(8) & 0.0006(1)\\
Te10 & 0.0062(1) & 0.0066(1) & 0.0058(1) & 0.000(1) & -0.00027(8) & -0.0002(1)\\
Te11 & 0.0078(1) & 0.0062(1) & 0.0066(1) & -0.0002(1) & 0.00045(8) & 0.0001(1)\\
Te12 & 0.0068(1) & 0.0066(1) & 0.0065(1) & -0.0002(1) & -0.00033(8) & -0.0004(1)\\\hline
Co1 & 0.0058(3) & 0.0044(2) & 0.0059(3) & -0.0002(2) & -0.0004(2) & 0.0000(2)\\
Co2 & 0.0058(3) & 0.0043(2) & 0.0063(3) & 0.0001(2) & -0.0008(2) & 0.0002(2)\\
Co3 & 0.0067(3) & 0.0043(2) & 0.0070(3) & -0.0007(2) & -0.0004(2) & 0.0002(2)\\
Co4 & 0.0068(3) & 0.0042(2) & 0.0074(3) & 0.0000(2) & 0.0002(2) & 0.0000(2)\\
Co5 & 0.0074(3) & 0.0042(2) & 0.0068(3) & 0.0003(2) & 0.0002(2) & -0.0005(2)\\
Co6 & 0.0060(3) & 0.0045(2) & 0.0074(3) & -0.0003(2) & -0.0004(2) & -0.0002(2)\\
Co7 & 0.0069(3) & 0.0042(2) & 0.0067(3) & -0.0006(2) & 0.0000(2) & -0.0004(2)\\
Co8 & 0.0055(3) & 0.0043(2) & 0.0076(3) & -0.0002(2) & -0.0003(2) & 0.0000(2)\\\hline
K1 & 0.0169(4) & 0.0178(5) & 0.0172(5) & -0.0007(4) & -0.0055(4) & 0.0021(4)\\
K2 & 0.0113(4) & 0.0245(5) & 0.0196(5) & 0.0015(4) & 0.0032(3) & 0.0021(4)\\
K3 & 0.0240(5) & 0.0173(5) & 0.0129(5) & -0.0009(4) & 0.0037(4) & -0.0008(4)\\
K4 & 0.0228(5) & 0.0182(5) & 0.0125(4) & -0.0012(4) & -0.0011(4) & -0.0016(4)\\
K5 & 0.0204(5) & 0.0299(6) & 0.0137(5) & 0.0049(4) & 0.0026(4) & -0.0024(4)\\
K6 & 0.0133(4) & 0.0240(5) & 0.0220(5) & -0.0018(4) & -0.0029(4) & 0.0002(4)\\
K7 & 0.0235(5) & 0.0415(7) & 0.0185(6) & 0.0050(5) & -0.0084(4) & -0.0110(5)\\
K8 & 0.0127(6) & 0.047(1) & 0.025(1) & -0.0149(8) & 0.0007(6) & 0.0030(6)\\
K8' & 0.015(3) & 0.045(4) & 0.025(5) & -0.008(4) & 0.005(3) & 0.001(3) \\\hline   
O1 & 0.005(1) & 0.010(1) & 0.013(1) & 0.002(1) & 0.002(1) & 0.003(1)\\
O2 & 0.010(2) & 0.011(1) & 0.012(2) & -0.002(1) & 0.000(1) & -0.006(1)\\
O3 & 0.010(2) & 0.010(1) & 0.009(2) & -0.004(1) & -0.002(1) & -0.002(1)\\
O4 & 0.011(2) & 0.010(1) & 0.013(2) & 0.004(1) & 0.001(1) & 0.001(1)\\
O5 & 0.015(2) & 0.010(1) & 0.014(2) & 0.002(1) & -0.001(1) & 0.006(1)\\
O6 & 0.006(1) & 0.010(1) & 0.011(1) & 0.000(1) & 0.002(1) & 0.003(1)\\
O7 & 0.013(2) & 0.011(1) & 0.012(2) & -0.005(1) & -0.002(1) & 0.004(1)\\
O8 & 0.016(2) & 0.011(1) & 0.009(2) & -0.004(1) & 0.000(1) & 0.007(1)\\
O9 & 0.011(2) & 0.009(1) & 0.010(2) & 0.000(1) & -0.002(1) & 0.005(1)\\
O10 & 0.008(1) & 0.012(1) & 0.015(2) & -0.001(1) & -0.001(1) & -0.004(1)\\
O11 & 0.011(1) & 0.007(1) & 0.007(1) & 0.001(1) & 0.002(1) & 0.002(1)\\\hline
    \end{tabular} 
\end{table}

\begin{table}[ht]
    \centering
    \caption{\textbf{K$_2$Co$_2$(TeO$_{3}$)$_{3}$~$\cdot$~2.5~D$_2$O  single crystal structural refinement data - anisotropic displacement parameters}. All $U_{ij}$ values are given in units of $\textrm\AA{^2}$.}
    \vspace{1mm}
    \label{SC_U_2}
    \renewcommand{\arraystretch}{0.6}
   \begin{tabular}{|c|c|c|c|c|c|c|}
   \hline
   Atom & $U_{11}$ & $U_{22}$ & $U_{33}$ & $U_{12}$ & $U_{13}$ & $U_{23}$\\\hline
O12 & 0.009(1) & 0.006(1) & 0.006(1) & -0.001(1) & -0.001(1) & 0.001(1)\\
O13 & 0.012(2) & 0.011(1) & 0.011(2) & 0.000(1) & 0.000(1) & -0.005(1)\\
O14 & 0.010(2) & 0.007(1) & 0.013(2) & 0.004(1) & 0.000(1) & -0.003(1)\\
O15 & 0.010(1) & 0.008(1) & 0.006(1) & 0.001(1) & 0.000(1) & 0.001(1)\\
O16 & 0.010(1) & 0.007(1) & 0.008(1) & -0.002(1) & 0.000(1) & -0.002(1)\\
O17 & 0.013(2) & 0.010(1) & 0.010(2) & -0.003(1) & -0.001(1) & 0.004(1)\\
O18 & 0.010(1) & 0.012(1) & 0.012(2) & 0.005(1) & -0.001(1) & -0.003(1)\\
O19 & 0.016(2) & 0.012(2) & 0.009(2) & 0.002(1) & -0.001(1) & 0.007(1)\\
O20 & 0.014(2) & 0.008(1) & 0.011(2) & 0.001(1) & -0.002(1) & 0.005(1)\\
O21 & 0.011(2) & 0.013(2) & 0.017(2) & 0.006(1) & -0.002(1) & -0.004(1)\\
O22 & 0.009(1) & 0.006(1) & 0.006(1) & 0.001(1) & -0.003(1) & 0.000(1)\\
O23 & 0.008(1) & 0.006(1) & 0.007(1) & 0.003(1) & 0.000(1) & 0.002(1)\\
O24 & 0.011(2) & 0.010(1) & 0.012(2) & 0.004(1) & 0.000(1) & 0.001(1)\\
O25 & 0.009(1) & 0.006(1) & 0.009(1) & -0.000(1) & -0.002(1) & 0.001(1)\\
O26 & 0.007(1) & 0.009(1) & 0.009(1) & -0.001(1) & -0.003(1) & 0.003(1)\\
O27 & 0.011(2) & 0.011(1) & 0.010(2) & -0.004(1) & -0.001(1) & 0.004(1)\\
O28 & 0.014(2) & 0.011(2) & 0.010(2) & -0.007(1) & -0.001(1) & -0.001(1)\\
O29 & 0.015(2) & 0.009(1) & 0.013(2) & 0.006(1) & 0.000(1) & 0.000(1)\\
O30 & 0.008(1) & 0.006(1) & 0.008(1) & 0.002(1) & 0.001(1) & 0.002(1)\\
O31 & 0.012(2) & 0.010(1) & 0.011(2) & 0.005(1) & 0.004(1) & 0.002(1)\\
O32 & 0.009(2) & 0.013(2) & 0.014(2) & -0.007(1) & 0.000(1) & -0.001(1)\\
O33 & 0.007(1) & 0.008(1) & 0.011(1) & 0.001(1) & -0.003(1) & 0.002(1)\\
O34 & 0.013(2) & 0.018(2) & 0.023(2) & -0.001(1) & -0.002(1) & -0.001(1)\\
O35 & 0.012(2) & 0.013(2) & 0.016(2) & -0.008(1) & -0.002(1) & -0.002(1)\\
O36 & 0.019(2) & 0.021(2) & 0.012(2) & 0.005(1) & -0.004(1) & -0.001(1)\\
O37 & 0.012(2) & 0.011(2) & 0.012(2) & -0.001(1) & -0.001(1) & -0.005(1)\\
O38 & 0.010(2) & 0.010(2) & 0.013(2) & 0.004(1) & -0.001(1) & -0.004(1)\\
O39 & 0.014(1) & 0.025(2) & 0.023(2) & 0.005(1) & -0.001(1) & 0.002(1)\\
O40 & 0.011(1) & 0.031(2) & 0.020(2) & -0.005(2) & -0.002(1) & 0.002(1)\\
O41 & 0.019(2) & 0.026(2) & 0.022(2) & -0.003(2) & 0.004(1) & -0.006(1)\\
O42 & 0.025(2) & 0.044(2) & 0.019(2) & 0.005(2) & 0.004(2) & -0.013(2)\\
O43 & 0.015(2) & 0.043(2) & 0.019(2) & -0.015(2) & -0.004(1) & 0.002(2)\\
O44 & 0.020(2) & 0.032(2) & 0.013(2) & 0.001(2) & 0.003(1) & -0.001(2)\\
O45 & 0.020(2) & 0.045(2) & 0.014(2) & 0.001(2) & 0.004(1) & 0.004(2)\\
O46 & 0.020(2) & 0.026(2) & 0.013(2) & -0.000(2) & -0.003(2) & 0.004(2)\\
O46' & 0.011(7) & 0.029(7) & 0.022(8) & 0.001(7) & -0.012(6) & -0.005(6)\\\hline
    \end{tabular} 
\end{table}

\begin{table}[ht]
\centering
\caption{\textbf{K$_2$Zn$_2$(TeO$_{3}$)$_{3}$~$\cdot$~1.5~(H$_2$O) single crystal structural refinement data - atomic positions and site occupancies}}
\vspace{1mm}
\label{SC_Zn_1}
\renewcommand{\arraystretch}{0.8}
\begin{tabular}{|c|c|c|c|c|}
    \hline
    \textbf{Atom} & \textbf{x} & \textbf{y} & \textbf{z} & \textbf{Occ.} \\ \hline
    Te1 & 0.50810(6) & 0.46866(6) & $\frac{1}{4}$ & 1 \\
    Zn1 & $\frac{2}{3}$ & $\frac{1}{3}$ & 0.5623(2) & 1 \\
    O1 & 0.6611(5) & 0.5204(5) & 0.4299(5) & 1 \\
    O2 & 0.5060(6) & 0.6669(6) & $\frac{1}{4}$ & 1 \\\hline
    \end{tabular}
\end{table}

\begin{table}[ht]
    \centering
    \caption{\textbf{K$_2$Zn$_2$(TeO$_{3}$)$_{3}$~$\cdot$~1.5~(H$_2$O) single crystal structural refinement data - anisotropic displacement parameters}. All $U_{ij}$ values are given in units of $\textrm\AA{^2}$.}
    \vspace{1mm}
    \label{SC_Zn_2}
    \renewcommand{\arraystretch}{0.8}
   \begin{tabular}{|c|c|c|c|c|c|c|}
   \hline
   Atom & \textbf{$U_{11}$} & \textbf{$U_{22}$} & \textbf{$U_{33}$} & \textbf{$U_{12}$} & \textbf{$U_{13}$} & \textbf{$U_{23}$}\\\hline
   Te1 & 0.0213(3) & 0.0195(3) & 0.0197(3) & 0 & 0 & 0.0122(2) \\
   Zn1 & 0.0181(4) & 0.0181(4) & 0.0097(5) & 0 & 0 & 0.0091(2) \\
   O1 & 0.029(2) & 0.024(2) & 0.028(2) & 0.004(2) & -0.007(2) & 0.013(2) \\
   O2 & 0.018(2) & 0.020(2) & 0.017(2) & 0 & 0 & 0.012(2) \\\hline
    \end{tabular} 
\end{table}

\begin{figure}[ht]
    \centering
    \includegraphics[width=0.7\linewidth]{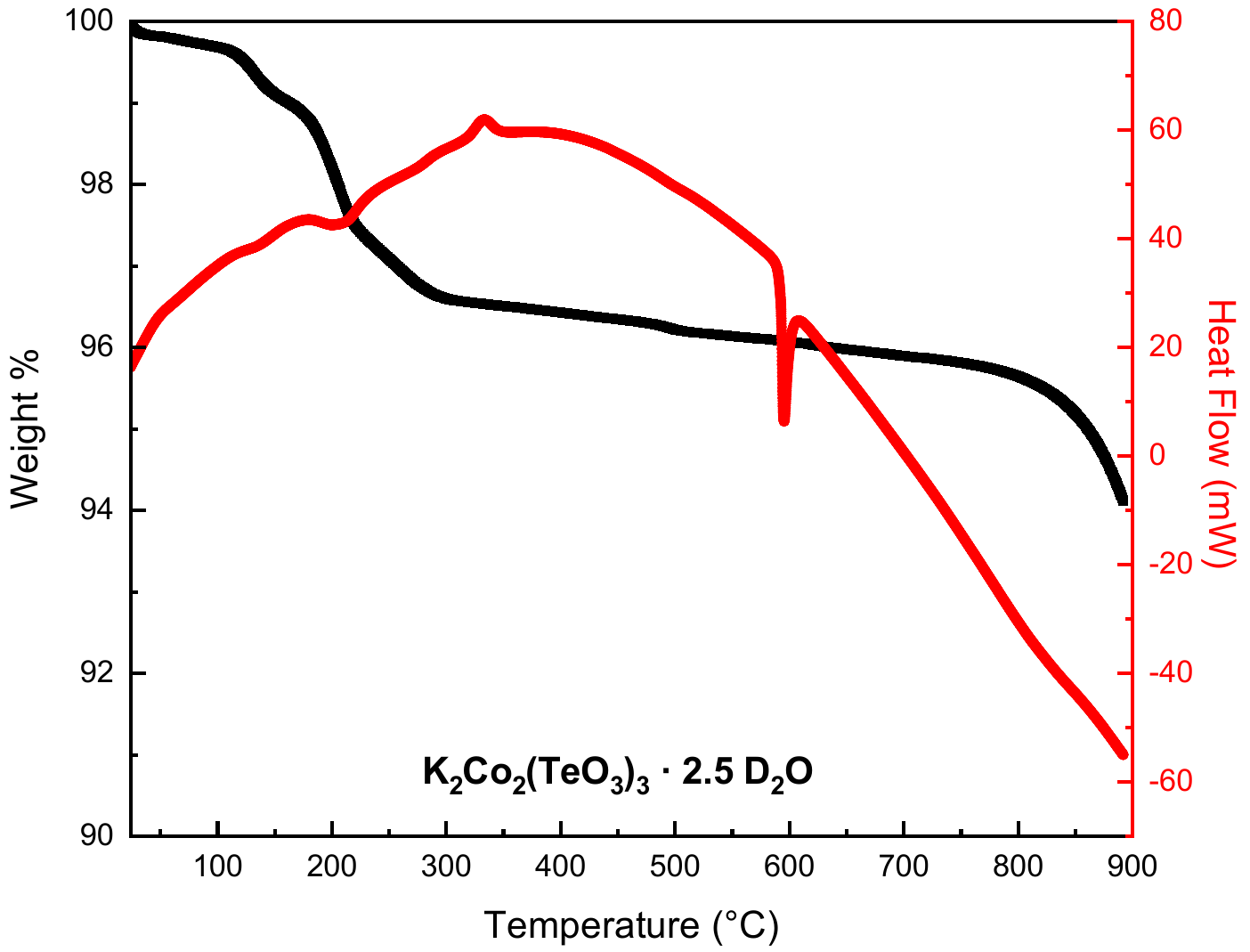}
    \vspace{-5mm}
    \caption{\textbf{Thermal evolution of KCoTOD.} Percent change in sample mass (black) and heat flow (red) as a function of temperature for as-synthesized KCoTOD samples.}
    \label{TGADTA}
\end{figure}
\clearpage

\subsection*{Magnetic characterization}
\begin{table}[ht]
    \centering
    \caption{\textbf{Fitting parameters obtained from analysis of oriented KCoTOH single crystal magnetization measurements from \emph{T}~=~70~-~300~K.}}
    \vspace{1mm}
    \label{Oriented_CW}
    \begin{tabular}{|c|c|c|c|c|}
    \hline
     \textbf{Orientation} & \textbf{\emph{T}$_\text{N}$ (K)} & \textbf{$\theta_{\text{CW}}$ (K)} & \textbf{$p_{\text{eff}}$ ($\mu_\text{B}$/Co)} &  \textbf{$\chi_\text{o}$} \\\hline
    $\mu_\text{o}H~//~$~c & 7.6 & -73.0(1) & 4.50(1) & -4.49(4)~x~10$^{-3}$\\\hline
    $\mu_\text{o}H\perp$~c & 7.1 & -65.5(1) & 4.70(1) & -6.93(4)~x~10$^{-3}$\\\hline
    \end{tabular}
\end{table}

\begin{figure}[ht]
    \centering
    \includegraphics[width=1.0\linewidth]{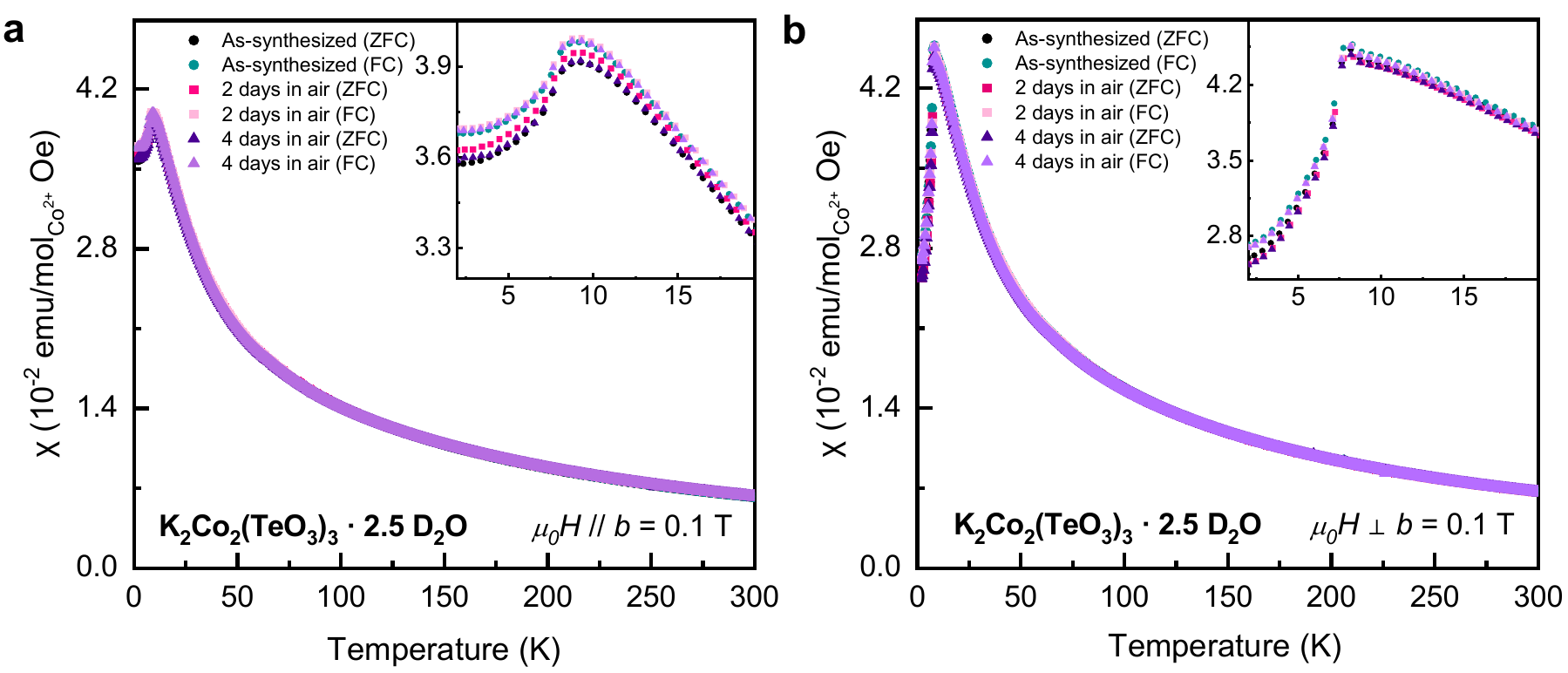}
    \vspace{-10mm}
    \caption{\textbf{Evidence for minimal additional hydration in KCoTOD upon prolonged exposure to air.} a) Magnetic susceptibility as a function of temperature for representative as-synthesized (green) KCoTOD single crystals oriented with a) $\mu_\text{o}H~//~b$ and b) $\mu_\text{o}H\perp~b$, as well as after 2 days (pink) and 4 days (purple) of exposure to ambient air.}
    \label{Hydrated_mag}
\end{figure}


\begin{figure}[ht]
    \centering
    \includegraphics[width=0.7\linewidth]{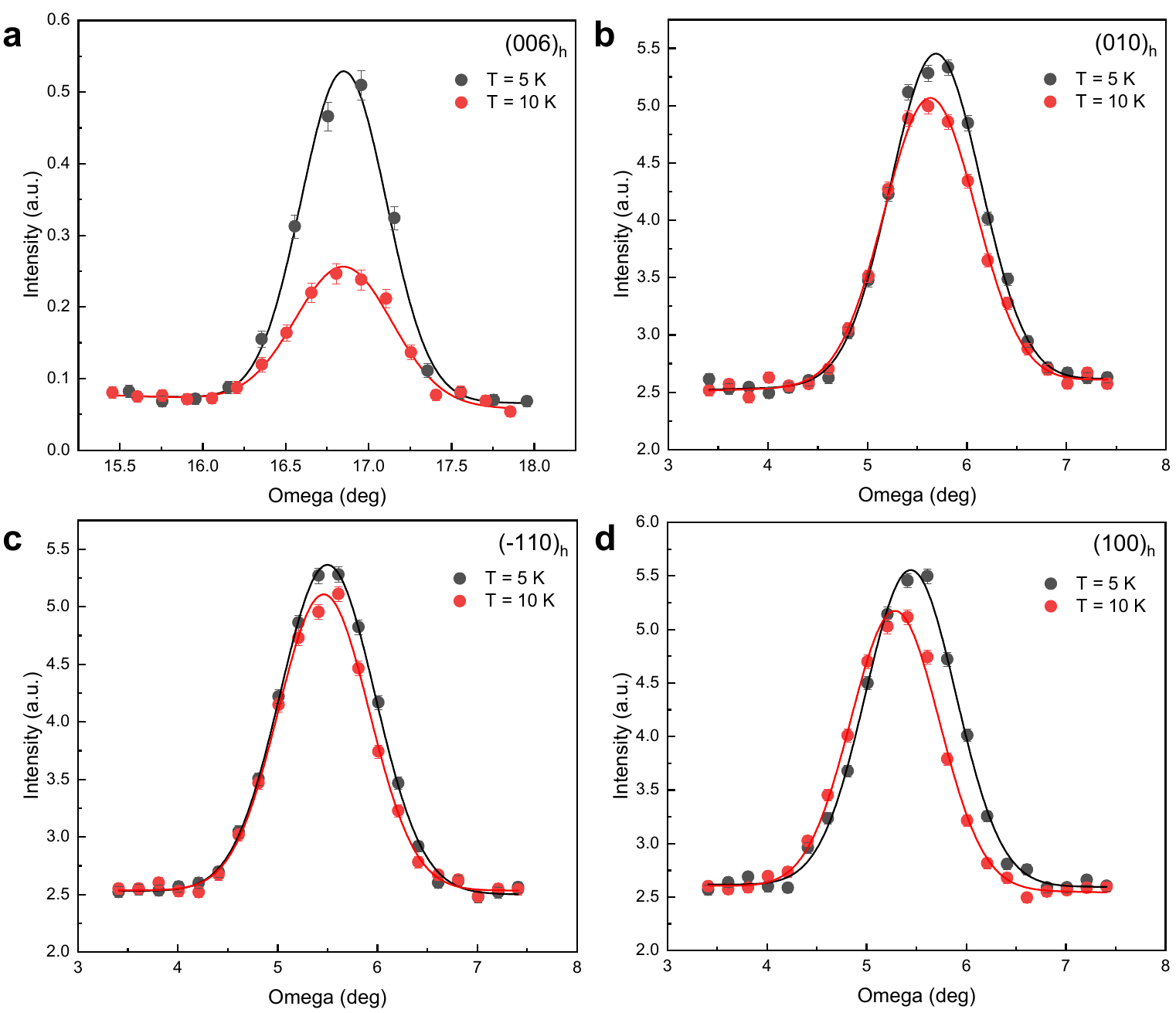}
    \caption{\textbf{Rocking curve scans} of the a) (006)$_\text{h}$, b) (010)$_\text{h}$, c) (-110)$_\text{h}$, and d) (100)$_\text{h}$ reflections at \textit{T}~=~5~K (black) and \textit{T}~=~10~K (red) of the KCoTOH structure in the hexagonal $P6_3/m$ (\#176) space group. The solid lines represent sloped Gaussian fits to the data. The slight shift to lower angles observed between the two temperature datasets for b)~-~d) is likely due to a shifting of the sample during the measurement, as a) was collected separately. The increase in intensity for all reflections below $T_\text{N}$ supports $k~=~0$ magnetic ordering.}
    \label{Neutron2}
\end{figure}
\clearpage

\begin{table}[ht]
    \centering
    \caption{\textbf{All possible commensurate magnetic subgroups of the nuclear $P6_3/m$ (\#176) space group.}}
    \vspace{1mm}
    \label{Irrep_table}
        \begin{tabular}{|c|c|c|c|c|c|}
    \hline
     \textbf{Subgroup} & \textbf{\#} & \textbf{Atomic coord.} & \textbf{Moment} & \textbf{FM/AFM} & $\mathbf{\chi^2}$ \\\hline
    $P6{_3}'/m'$ & 176.147 & (1/3, 2/3, z) & (0, 0, $m_z$) & AFM & - \\\hline
    $P6{_3}/m'$ & 176.146 & (1/3, 2/3, z) & (0, 0, $m_z$) & AFM & - \\\hline
    $P6{_3}'/m$ & 176.145 & (1/3, 2/3, z) & (0, 0, $m_z$) & AFM & - \\\hline
    $P6{_3}/m$ & 176.143 & (1/3, 2/3, z) & (0, 0, $m_z$) & FM & - \\\hline
    $P\overline{6}'$ & 174.135 & (1/3, 2/3, z); (2/3, 1/3, z-1/2) & (0, 0, $m_z$) & AFM & - \\\hline
    $P\overline{6}$ & 174.133 & (1/3, 2/3, z); (2/3, 1/3, z-1/2) & (0, 0, $m_z$) & FM & - \\\hline
    $P6{_3}'$ & 173.131 & (1/3, 2/3, z); (2/3, 1/3, z-1) & (0, 0, $m_z$) & AFM & - \\\hline
    $P6_3$ & 173.129 & (1/3, 2/3, z); (2/3, 1/3, z-1) & (0, 0, $m_z$) & FM & - \\\hline
    $P\overline{3}'$ & 147.15 & (1/3, 2/3, z); (2/3, 1/3, z-1/2) & (0, 0, $m_z$) & AFM & - \\\hline
    $P\overline{3}$ & 147.13 & (1/3, 2/3, z); (2/3, 1/3, z-1/2) & (0, 0, $m_z$) & FM & - \\\hline
    $P3$ & 143.1 & (1/3, 2/3, z); (2/3, 1/3, z-1/2) & (0, 0, $m_z$) & FM & - \\
    & & (2/3, 1/3, z-1); (1/3, 2/3, z-3/2) & & & \\\hline
    $P2{_1}'/m'$ & 11.54 & (1/3, 2/3, z) & ($m_x$, $m_y$, $m_z$) & AFM & 18.7 \\\hline
    $P2{_1}/m'$ & 11.53 & (1/3, 2/3, z) & ($m_x$, $m_y$, $m_z$) & AFM & 1.23 \\\hline
    $P2{_1}'/m$ & 11.52 & (1/3, 2/3, z) & ($m_x$, $m_y$, $m_z$) & AFM & 272.0 \\\hline
    $P2{_1}/m$ & 11.50 & (1/3, 2/3, z) & ($m_x$, $m_y$, $m_z$) & FM & - \\\hline
    $Pm$' & 6.20 & (1/3, 2/3, z); (2/3, 1/3, z-1/2) & ($m_x$, $m_y$, $m_z$) & AFM & 1.51 \\\hline
    $Pm$ & 6.18 & (1/3, 2/3, z); (2/3, 1/3, z-1/2) & ($m_x$, $m_y$, $m_z$) & FM & - \\\hline
    $P2{_1}'$ & 4.9 & (1/3, 2/3, z); (2/3, 1/3, z-1) & ($m_x$, $m_y$, $m_z$) & AFM & 9.02 \\\hline
    $P2{_1}$ & 4.7 & (1/3, 2/3, z); (2/3, 1/3, z-1) & ($m_x$, $m_y$, $m_z$) & FM & - \\\hline
    $P\overline{1}'$ & 2.6 & (1/3, 2/3, z); (2/3, 1/3, z-1/2) & ($m_x$, $m_y$, $m_z$) & AFM & 1.19 \\\hline
    $P\overline{1}$ & 2.4 & (1/3, 2/3, z); (2/3, 1/3, z-1/2) & ($m_x$, $m_y$, $m_z$) & FM & - \\\hline
    $P1$ & 1.1 & (1/3, 2/3, z); (2/3, 1/3, z-1/2) & ($m_x$, $m_y$, $m_z$) & FM & - \\
    & & (2/3, 1/3, z-1); (1/3, 2/3, z-3/2) & & & \\\hline
    \end{tabular}
\end{table}

\subsection*{$\mu$SR analysis}
\begin{figure}[ht]
    \includegraphics[width=0.8\textwidth]{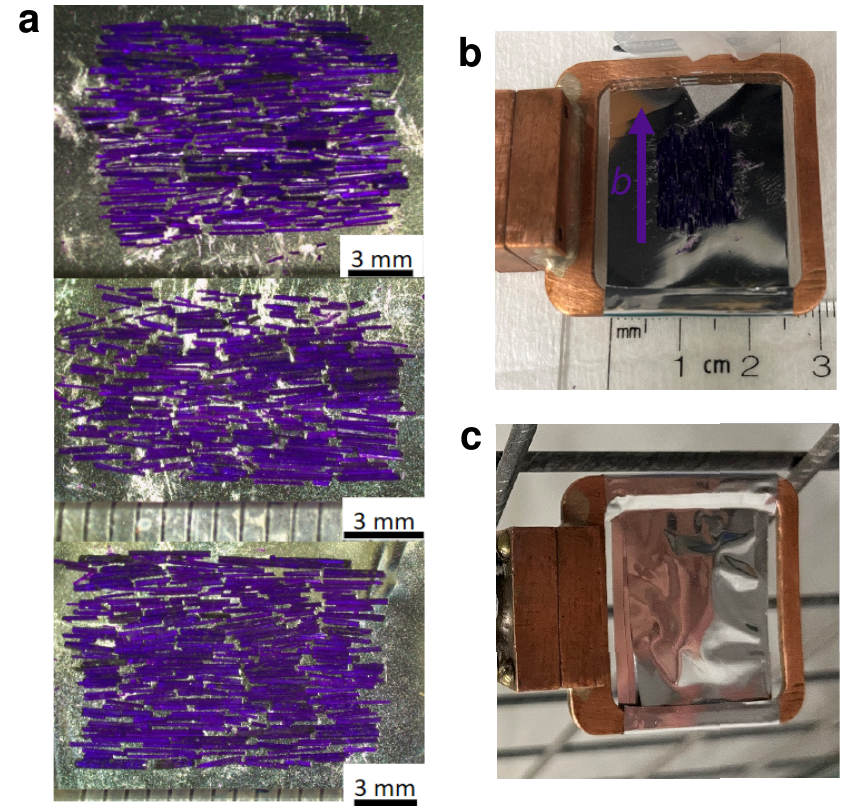}
    \caption{\textbf{Experimental setup for the muon spin relaxation measurements reported in this work.} a) KCoTOH single crystals were roughly coaligned on three strips of silver tape, which were then b) mounted to the sample holder and c) carefully stacked.}
    \label{muSR_exp_setup}
\end{figure}

\begin{table}[ht]
    \centering
    \caption{\textbf{Fitting parameters obtained from analysis of zero-field muon spin relaxation measurements performed on a mosaic of coaligned KCoTOH single crystals in the \textbf{spin-rotated geometry} ($p_{\mu}~\perp~b$).} All spectra were fit by Equation 1 with the following parameters held globally constant: $\alpha$~=~0.9749, A$_1$~=~0.1134, A$_2$~=~0.0612, A$_3$~=~0.0128, $\varphi_1$ = 0, $\varphi_2$ = 0, $\varphi_3$ = 0. Approximately 79\% of the total asymmetry observed in this orientation was attributed to the sample.}
    \vspace{1mm}
    \label{muSR_params_ZF}
    \begin{tabular}{|c|c|c|c|c|c|c|c|}
    \hline
     \textbf{Temperature} & $\mathbf{\sigma_1}$ & $\mathbf{\sigma_2}$ & $\mathbf{\sigma_3}$ & $\mathbf{\nu_1}$ &  $\mathbf{\nu_2}$ & $\mathbf{\nu_3}$ & $\mathbf{\chi^2}$ \\[-3pt]
     (K) & ($\mu$s$^{-1}$) & ($\mu$s$^{-1}$) & ($\mu$s$^{-1}$) & (T) & (T) & (T) & \\\hline
    8.099(9) & 18(1) & 2(1) & 3(3) & 0 & 0 & 0 & 1.05 \\\hline
    7.999(7) & 11(1) & 25(1) & 7(1) & 0 & 0.020(1) & 0.030(1) & 1.06 \\\hline
    7.899(8) & 21(4) & 17(4) & 11(4) & 0 & 0.032(8) & 0.060(4) & 1.03 \\\hline
    7.80(1) & 23(1) & 15(3) & 12(5) & 0 & 0.046(2) & 0.069(9) & 0.99 \\\hline
    7.56(1) & 25(1) & 15(1) & 9(2) & 0.020(3) & 0.067(1) & 0.093(3) & 1.08 \\\hline
    7.40(1) & 32(4) & 18(1) & 10(2) & 0.032(3) & 0.084(2) & 0.120(3) & 1.06 \\\hline
    7.20(1) & 42(5) & 21(1) & 19(5) & 0.041(4) & 0.098(3) & 0.130(1) & 0.99 \\\hline
    7.00(1) & 44(3) & 22(1) & 20(3) & 0.048(2) & 0.113(1) & 0.160(1) & 0.98 \\\hline
    6.80(2) & 51(6) & 24(1) & 17(4) & 0.059(3) & 0.121(4) & 0.188(7) & 1.05 \\\hline
    6.40(2) & 55(5) & 23(1) & 20(4) & 0.077(3) & 0.135(3) & 0.195(9) & 0.96 \\\hline
    6.06(2) & 59(4) & 24(1) & 13(2) & 0.087(4) & 0.148(3) & 0.211(4) & 1.00 \\\hline
    5.79(3) & 60(4) & 25(1) & 11(2) & 0.101(5) & 0.158(3) & 0.224(3) & 1.08 \\\hline
    4.99(3) & 72(6) & 26(1) & 10(1) & 0.114(7) & 0.176(3) & 0.241(2) & 1.08 \\\hline
    4.20(2) & 85(13) & 26(1) & 9(2) & 0.124(8) & 0.180(2) & 0.250(2) & 1.09 \\\hline
    3.39(5) & 90(3) & 27(1) & 10(2) & 0.134(5) & 0.186(2) & 0.254(1) & 1.05 \\\hline
    2.54(4) & 92(3) & 28(1) & 10(2) & 0.14(1) & 0.192(2) & 0.258(2) & 1.07 \\\hline
    1.897(1) & 95(7) & 26(1) & 8(1) & 0.15(1) & 0.192(2) & 0.259(1) & 1.02 \\\hline
    \end{tabular}
\end{table}

\begin{table}[ht]
    \centering
    \caption{\textbf{Fitting parameters obtained from analysis of zero-field muon spin relaxation measurements performed on a mosaic of coaligned KCoTOH single crystals in the \textbf{non-spin-rotated geometry} ($p_{\mu}~//~b$)}. All spectra were fit with Equation 2 with the following parameters held globally constant: $\alpha$~=~1.2022, A~=~0.08824, $\varphi$ = 0. Approximately 46\% of the total asymmetry observed in this orientation was attributed to the sample.}
    \vspace{1mm}
    \label{muSR_params_NSR}
        \begin{tabular}{|c|c|c|c|}
    \hline
     \textbf{Temperature (K)} & $\mathbf{\sigma}$ ($\mu$s$^{-1}$) & $\mathbf{\nu}$ (T) & $\mathbf{\chi^2}$ \\\hline
    8.25(2) & 14(1) & 0 & 1.075 \\\hline
    8.00(2) & 16(1) & 0 & 1.145 \\\hline
    7.76(2) & 26(2) & 0.033(1) & 1.048 \\\hline
    7.53(2) & 32(2) & 0.047(2) & 1.075 \\\hline
    6.49(2) & 43(1) & 0.119(5) & 1.045 \\\hline
    5.25(2) & 46(3) & 0.163(5) & 1.049 \\\hline
    3.6(1) & 50(3) & 0.185(5) & 1.028 \\\hline
    2.063(7) & 50(2) & 0.195(3) & 1.064 \\\hline
    1.883(1) & 53(2) & 0.195(2) & 1.048 \\\hline
    \end{tabular}
\end{table}

\begin{figure}[ht]
    \centering
    \includegraphics[width=0.8\textwidth]{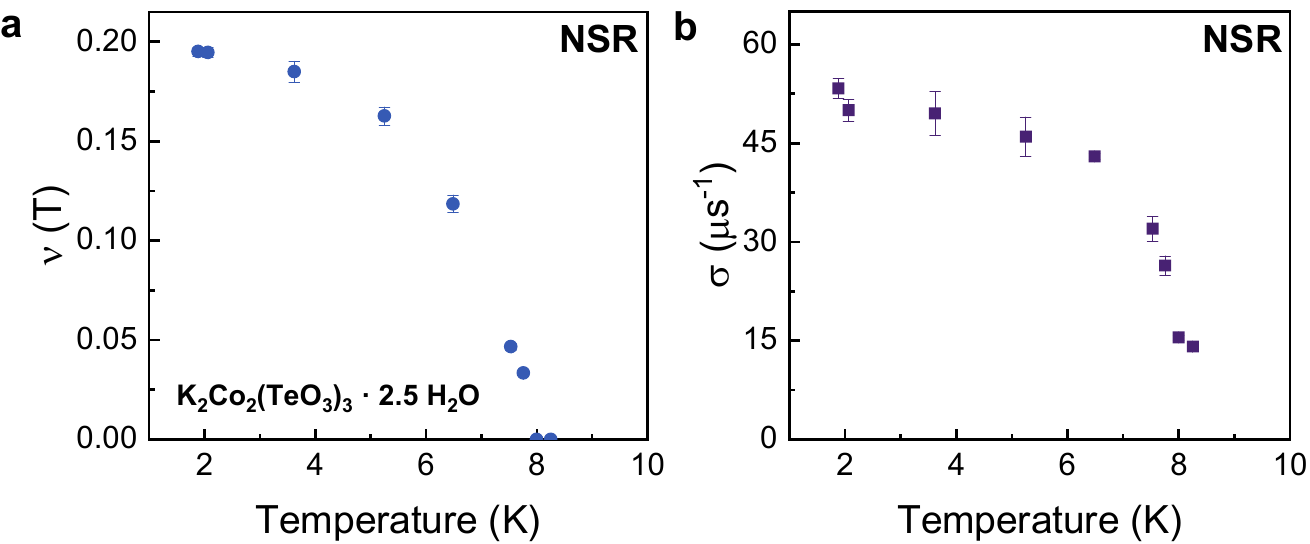}
    \caption{\textbf{Zero-field muon spin relaxation measurement-derived a) internal field strength and b) exponential decay rate of a mosaic of coaligned KCoTOH single crystals in the \textbf{non-spin-rotated geometry} ($p_{\mu}~//~b$).}}
    \label{wTF_NSR_ZF}
\end{figure}

\begin{figure}[ht]
    \centering
    \includegraphics[width=0.9\textwidth]{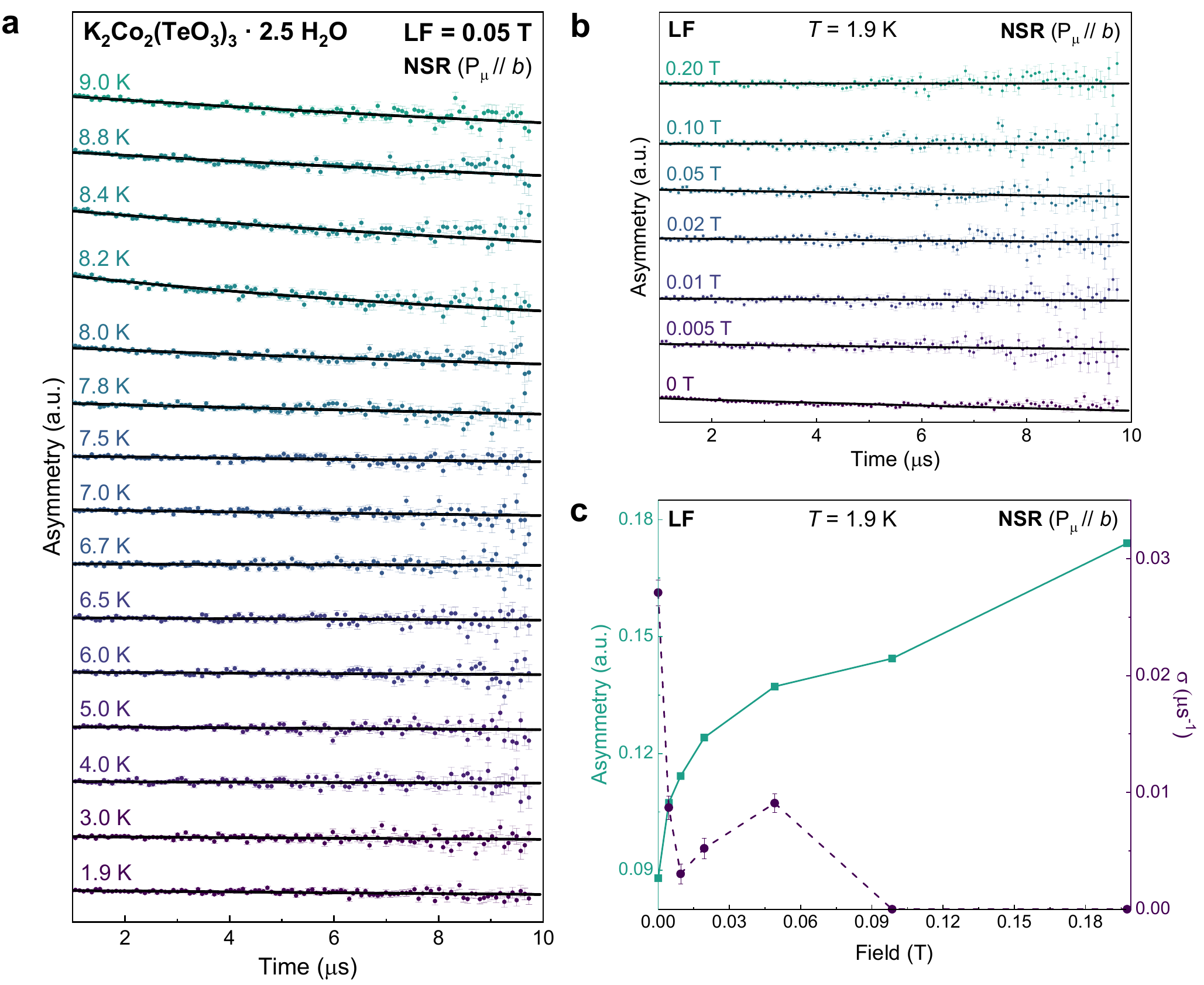}
    \caption{\textbf{a) Temperature-dependent longitudinal field (LF = 0.05 T) and b) isothermal, LF-dependent muon spin relaxation measurements at \textit{T}~=~1.9~K collected in the \textbf{non-spin-rotated geometry} ($p_{\mu}~//~b$).} c) Sample asymmetry (green) and relaxation rate (purple) as a function of longitudinal field strength.}
    \label{LF_fig}
\end{figure}

\begin{table}[ht]
    \centering
    \caption{\textbf{Fitting parameters obtained from analysis of field-dependent longitudinal field muon spin relaxation measurements performed on a mosaic of coaligned KCoTOH single crystals in the \textbf{non-spin-rotated geometry} ($p_{\mu}~//~b$) at \textit{T}~=~1.88(1)~K.} All spectra were fit with Equation S4 from \textit{t}~=~1~-~10~$\mu$s with the following parameters held globally constant: $\alpha$~=~1.2022, $\varphi$ = 0.}
    \vspace{1mm}
    \label{muSR_params_NSR_LF_B}
        \begin{tabular}{|c|c|c|c|}
    \hline
     \textbf{Field (T)} & \textbf{Asymmetry (a.u.)} & $\mathbf{\sigma}$ ($\mu$s$^{-1}$) & \textbf{Reduced} $\mathbf{\chi^2}$ \\\hline
    0 & 0.0880(3) & 0.027(1) & 2.63 \\\hline
    0.005 & 0.1073(4) & 0.009(1) & 1.11 \\\hline
    0.010 & 0.1142(3) & 0.003(1) & 1.02 \\\hline
    0.019 & 0.1241(4) & 0.005(1) & 1.16 \\\hline
    0.049 & 0.1372(4) & 0.009(1) & 1.15 \\\hline
    0.099 & 0.1444(4) & 0 & 1.50 \\\hline
    0.198 & 0.1740(4) & 0 & 0.94 \\\hline
    \end{tabular}
\end{table}

\begin{table}[ht]
    \centering
    \caption{\textbf{Fitting parameters obtained from analysis of temperature-dependent longitudinal field (LF = 0.05 T) muon spin relaxation measurements performed on a mosaic of coaligned KCoTOH single crystals in the non-spin-rotated geometry ($p_{\mu}~//~b$).} All spectra were fit with Equation S4 from \textit{t}~=~1~-~10~$\mu$s with the following parameters held globally constant: $\alpha$~=~1.2022, $\varphi$ = 0.}
    \vspace{1mm}
    \label{muSR_params_NSR_LF_T}
        \begin{tabular}{|c|c|c|c|}
    \hline
     \textbf{Temperature (K)} & \textbf{Asymmetry (a.u.)} & $\mathbf{\sigma}$ ($\mu$s$^{-1}$) & \textbf{Reduced} $\mathbf{\chi^2}$ \\\hline
    1.9(1) & 0.1343(3) & 0.012(1) & 1.16 \\\hline
    3.0(1) & 0.135(1) & 0.006(2) & 1.19 \\\hline
    4.0(1) & 0.136(1) & 0.003(3) & 0.94 \\\hline
    5.0(1) & 0.137(1) & 0.007(2) & 1.16 \\\hline
    6.0(1) & 0.136(1) & 0.006(2) & 1.09 \\\hline
    6.5(1) & 0.136(1) & 0.005(2) & 1.09 \\\hline
    6.7(1) & 0.134(1) & 0.003(2) & 1.06 \\\hline
    7.0(1) & 0.134(1) & 0.014(2) & 1.19 \\\hline
    7.5(1) & 0.131(1) & 0.014(2) & 0.86 \\\hline
    7.8(1) & 0.128(1) & 0.030(3) & 1.12 \\\hline
    8.0(1) & 0.133(1) & 0.048(3) & 1.33 \\\hline
    8.2(1) & 0.190(2) & 0.089(3) & 1.35 \\\hline
    8.4(1) & 0.216(1) & 0.060(2) & 1.38 \\\hline
    8.8(1) & 0.223(1) & 0.040(2) & 1.15 \\\hline
    9.0(1) & 0.227(1) & 0.045(2) & 1.37 \\\hline
    9.5(1) & 0.229(1) & 0.039(2) & 0.94 \\\hline
    10.0(1) & 0.233(1) & 0.041(2) & 1.09 \\\hline
    12.0(1) & 0.232(1) & 0.031(2) & 0.87 \\\hline
    15.0(1) & 0.237(1) & 0.032(2) & 0.94 \\\hline
    20.5(2) & 0.237(1) & 0.029(2) & 1.08 \\\hline
    51(1) & 0.235(1) & 0.023(2) & 1.05 \\\hline
    74(4) & 0.235(1) & 0.021(1) & 0.94 \\\hline
    \end{tabular}
\end{table}

\begin{figure}[ht]
    \centering
    \includegraphics[width=0.7\textwidth]{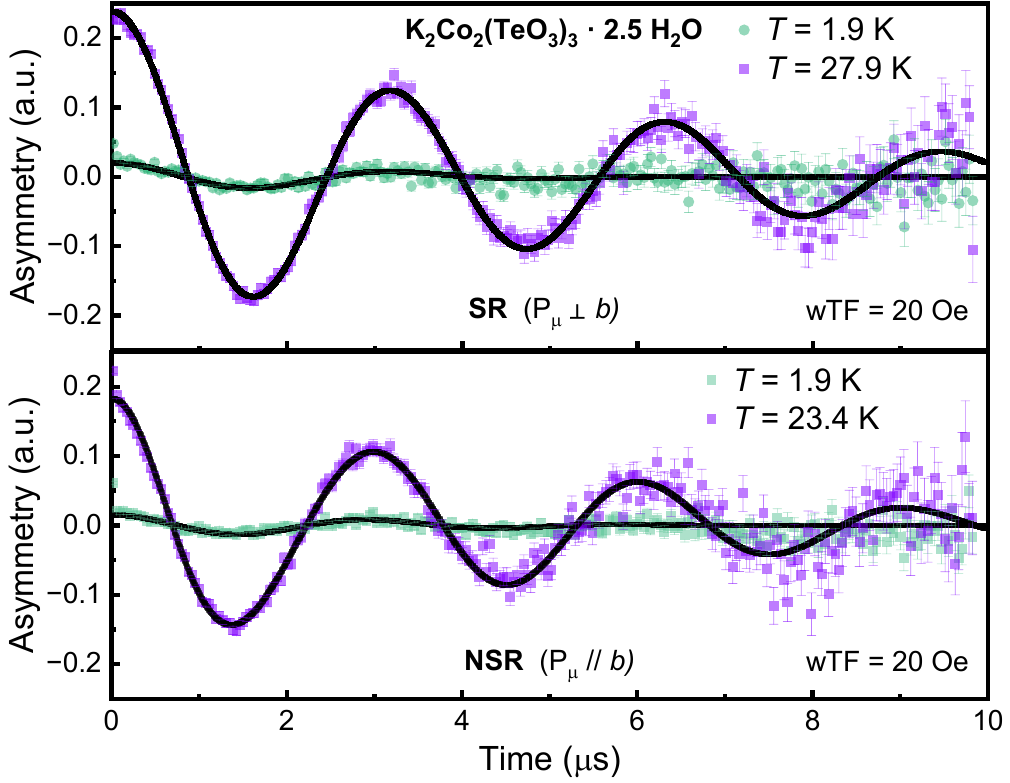}
    \vspace{-5mm}
    \caption{\textbf{Weak transverse field (wTF) muon spin relaxation spectra} collected from KCoTOH single crystals in the SR (top) and NSR (bottom) geometries, below (green) and above (purple) \textit{T}$_N$.}
    \label{wTF_NSR_SR_comb}
\end{figure}

\begin{table}[ht]
    \centering
    \caption{\textbf{Fitting parameters obtained from analysis of weak transverse field (wTF) muon spin relaxation measurements performed on a mosaic of coaligned KCoTOH single crystals in the non-spin-rotated (NSR) ($p_{\mu}~//~b$) and \textbf{spin-rotated (SR) ($p_{\mu}~\perp~b$)} geometries.}}
    \vspace{1mm}
    \label{wTF_fits}
    \begin{tabular}{|c|c|c|c|c|c|c|c|c|c|}
    \hline
    & Temp. & \textbf{$\alpha$} & \textbf{$A_1$} & \textbf{$A_2$} & $\mathbf{\sigma_1}$ & $\mathbf{\sigma_2}$ & $\mathbf{\nu_1}$ &  $\mathbf{\nu_2}$ & $\mathbf{\chi^2}$ \\[-3pt]
    & (K) & & & & ($\mu$s$^{-1}$) & ($\mu$s$^{-1}$) & (T) & (T) & \\\hline
    NSR & 1.9(1) & 1.036(1) & 0.016(1) & - & 0.4(1) & - & 0.00247(3) & - & 1.06\\\hline
    NSR & 23.4(1) & 1.202(2) & 0.129(5) & 0.054(5) & 0.2(1) & 0.8(1) & 0.00243(1) & 0.0032(1) & 1.10 \\\hline
    SR & 1.9(1) & 0.921(1) & 0.020(1) & - & 0.4(1) & - & 0.00223(4) & - & 1.08 \\\hline
    SR & 27.9(1) & 0.975(1) & 0.147(5) & 0.092(5) & 0.2(1) & 0.7(1) & 0.00232(1) & 0.0017(1) & 1.01 \\\hline
    \end{tabular}
\end{table}

\clearpage 



\end{document}